\documentclass[12pt]{article}
\usepackage{xcolor, soul}
\sethlcolor{yellow}
\usepackage{amsmath}
\usepackage{amssymb}
\usepackage{hyperref}
\begin{document}
\vskip 2cm
\begin{center}
{\bf {\Large Re-parameterization Invariance of  FRW Model: \vskip 0.2 cm   Supervariable and BRST Approaches}}\\

\vskip 1.2cm

{\sf B. Chauhan,    R. Tripathi  \\\vskip 0.4 cm

{\it  Department of Physics, Sunbeam Women's College Varuna,  \\Varanasi, 221 002 (Uttar Pradesh), India}}\\
\vskip 0.2cm

\vskip 0.3cm


\vskip 0.1cm

{\small {\sf {E-mails: bchauhan501@gmail.com; drraginitripathi09@gmail.com}}}

\end{center}

\vskip 0.5cm

\noindent
{\bf Abstract:}
We perform the Becchi-Rouet-Stora-Tyutin (BRST) quantization of a  \((0+1)\)-dimensional cosmological  Friedmann-Robertson-Walker (FRW) model. This quantization leverages the classical infinitesimal and continuous re-parameterization symmetry transformations of the system. To derive the nilpotent re-parameterization invariant BRST-anti-BRST symmetry transformations for the scale factor and corresponding momentum variables present in the cosmological FRW model, we employ the modified Bonora-Tonin supervariable approach (MBTSA) to BRST formalism. Through this approach, we also establish the BRST-anti-BRST invariant Curci-Ferrari (CF)-type restriction for this cosmological re-parameterization invariant model.
Further, we obtain the off-shell nilpotent quantum BRST-anti-BRST symmetry transformations for other variables within the model using the   (anti-)chiral supervariable approach (ACSA) to BRST formalism. Within the framework of ACSA, the CF-type restriction is demonstrated through two key aspects: $(i)$ the invariance of the coupled Lagrangians under symmetry transformations, and $(ii)$ the absolute anti-commutativity of the conserved BRST-anti-BRST charges. Notably, applying the MBTSA to a physical cosmological system, specifically a one-dimensional one, constitutes a novel contribution to this work. Additionally, in the application of ACSA, we restrict our analysis to   (anti-)chiral super expansions of supervariables, leading to the unique observation of the absolute anti-commutativity of the conserved BRST-anti-BRST charges.
Moreover, we highlight that the CF-type restriction demonstrates a universal nature, remaining consistent across any re-parameterization invariant models in general  D-dimensional spacetime.\\

\vskip 0.8 cm



\newpage 

\noindent
\section{Introduction}

\vskip 0.2 cm 

The foundational principles of local gauge theories play a pivotal role in providing an accurate theoretical framework for describing three out of four fundamental interactions of nature.
The covariant quantization of gauge field systems has a long-standing history, originating from the seminal works of Feynman [1], Faddeev and Popov [2], and DeWitt [3]. One of the methods of covariant quantization is the Becchi-Rouet-Stora-Tyutin (BRST) formalism [4-7] which stands out as an elegant and insightful method for quantizing local gauge theories. It ensures both unitarity and quantum gauge invariance are maintained at all orders of perturbative calculations for any physical process allowed by the local (interacting) gauge theory at the quantum level. A distinguishing aspect of the BRST formalism is the nilpotency of the BRST-anti-BRST symmetry transformations, alongside the absolute anti-commutativity between the BRST-anti-BRST transformations associated with a given local classical gauge symmetry.  One of the defining features of the BRST-anti-BRST symmetries is the presence of the BRST-anti-BRST invariant Curci-Ferrari (CF)-type restriction(s) [8, 9]. These restrictions not only ensure the absolute anti-commutativity of the BRST-anti-BRST symmetry transformations but also enable the existence of coupled but equivalent Lagrangians/Lagrangian densities for quantum gauge theories. In the special case of the Abelian 1-form gauge theory, the CF-type restriction becomes trivial, and the Lagrangian (or Lagrangian density) is uniquely defined. This serves as a limiting case of the non-Abelian 1-form gauge theory, where the CF-condition [10] plays a crucial role.
On the other hand, modern approaches to the covariant quantization of general gauge theories are grounded in either the BRST symmetry principle, as formulated in the renowned Batalin-Vilkovisky (BV) quantization framework [11], or the extended BRST symmetry principle, which underpins the quantization method developed by Batalin, Lavrov, and Tyutin (BLT) [12-14].

\vskip 0.15 cm

The usual superfield approach (USFA) to the BRST formalism [15-22] provides valuable insight into the geometrical foundation and interpretation of the off-shell nilpotency and absolute anti-commutativity properties of the BRST-anti-BRST symmetry transformations of the theory, where the horizontality condition (HC) plays a central and important role [17-19]. However, this method primarily focuses on deriving the BRST-anti-BRST symmetry transformations for the gauge field and the associated 
ghost-anti-ghost fields [17-19], leaving the BRST-anti-BRST symmetries of matter fields in interacting gauge field theories unaddressed.
In some previous research works (e.g., [23-26]), the USFA has been systematically extended and generalized. This extended approach incorporates the HC and leverages the gauge invariant restrictions (GIRs) to derive the BRST-anti-BRST symmetry transformations for the matter, (anti-)ghost, and gauge fields in interacting gauge field theories. Importantly, there is no inconsistency between the HC and GIRs, as they complement and enhance each other coherently and elegantly. This generalized framework has been termed the augmented version of the superfield approach (AVSA) to the BRST formalism.

\vskip 0.15 cm

In a recent series of papers [27-33], we introduced a simplified variant of the augmented version of the superfield/supervariable approach (AVSA), focusing exclusively on the (anti-)chiral supervariables/superfields and their relevant super expansions along the  (D, 1)-dimensional super sub-manifold of the general (D, 2)-dimensional supermanifold. This refined framework is called the  (anti-)chiral superfield/supervariable approach (ACSA). It is noteworthy that earlier superfield/supervariables approaches [15-26] involved full super expansions of superfields/supervariables along all Grassmannian directions of the general (D, 2)-dimensional supermanifold, applicable to D-dimensional local gauge invariant theories.
A defining feature of ACSA is its reliance on quantum gauge invariant restrictions [i.e., BRST-anti-BRST invariant restrictions] imposed on the supervariables/superfields. These restrictions facilitate the systematic derivation of BRST-anti-BRST symmetry transformations for all fields and variables in the (non-)interacting theory, along with the identification of BRST-anti-BRST invariant Curci-Ferrari (CF)-type restriction(s). A key outcome of ACSA is the observation that the conserved, nilpotent BRST-anti-BRST charges remain absolutely anti-commutative, even though the framework considers only the  (anti-)chiral super expansions of the supervariables/superfields with only one Grassmannian variable. This underscores the effectiveness and consistency of ACSA within the BRST formalism.

\vskip 0.15 cm

Incorporating the re-parameterization invariant symmetries (i.e. diffeomorphism invariant transformations) within the framework of the superfield approach to gauge theories has been a longstanding challenge for gauge invariant systems (see, e.g., [17-19]). This is a crucial step towards addressing gravitational and (super)string theories within the scope of the USFA/AVSA formalism. Recently, Bonora [34, 35] has achieved a significant advancement in this area by utilizing the superfield approach to derive both the appropriate BRST-anti-BRST transformations and the CF-type restriction for a D-dimensional re-parameterization invariant (i.e. diffeomorphism invariant) theory. We have termed this development as the modified Bonora-Tonin ({\cal B}T)   superfield approach (MBTSA) to BRST formalism.

In a recent couple of papers by our group [36-40], we have applied the theoretical beauty of the MBTSA as well as ACSA
[i.e.  (anti-)chiral superfield/supervariable approach] to BRST formalism to derive all symmetries. The re-parameterization invariant symmetries have been discussed for various theoretically and physically rich  1D re-parameterization invariant (1D diffeomorphism) systems such as free scalar relativistic particle [36], massive spinning relativistic particle [37], a free massive spinning relativistic particle (i.e. supersymmetric system) [38], non-relativistic free particle [39], scalar relativistic particle in interaction with an electromagnetic field [40]. We have also explored 2D re-parameterization (2D diffeomorphism) for a model of a Bosonic string theory in D-dimensional spacetime where
we obtained the 2D version of the universal CF-type restrictions [40] within the framework MBTSA.

\vskip 0.15 cm

The theory of general relativity (GR) is the most successful theory of gravitation, that describes the history of the universe and it is the basis of the modern cosmological models of the universe [41].  The FRW metric is called as standard model of modern cosmology. It is an exact solution of Einstein's field equation of GR. 
The re-parameterization invariant FRW model is a cosmological framework that maintains invariance under changes in the parametrization of time. It describes a homogeneous and isotropic expanding or contracting universe where the geometry is determined by the scale factor and which evolves according to the Friedmann equations derived from Einstein's field equations.
This invariance emphasizes that physical predictions of the model do not depend on the specific choice of the time parameter that reflects the general covariance of GR. It can be particularly useful when analyzing the dynamics of the universe in contexts like quantum cosmology or alternative formulations of the GR. Furthermore, 
the 1D re-parameterization invariant system holds substantial theoretical significance, serving as both a foundational model for (super)gauge invariant theories and a simplified representation of supergravity. Its broader generalization naturally leads to the formulation of superstring theory.
This system exemplifies the principle that continuous symmetries are fundamental to constructing meaningful physical theories.

\vskip 0.15 cm

The exploration of various quantum cosmological models is intriguing and challenging, particularly in the context of developing a quantum theory of gravity that unifies general relativity and quantum mechanics [42, 43]. The concept of homogeneous and isotropic spacetime symmetries was introduced through the  FRW models, which are widely recognized in cosmological studies [44-49]. These models have been fundamental in shaping modern cosmology, serving as the foundation for much of the research on quantum cosmology and dark energy in FRW spacetime.
However, it is important to note that many dark energy models encounter issues related to the cosmological constant (or dark energy density) such as fine-tuning and coincidence problems. As a result, a deeper examination of the underlying principles of cosmological FRW models in isotropic and homogeneous spacetimes is necessary. The BRST analysis has already been conducted for FRW models [50-54], this study is motivated by the desire to contribute significant and novel insights to the cosmological FRW model within the framework of BRST formalism. In the recent paper [30], the anti-BRST symmetry transformations for this cosmological model have been discussed in detail.

\vskip 0.15 cm

The present study aims to utilize the  (anti-)chiral supervariable approach (ACSA) to BRST formalism for analyzing   (0 + 1)-dimensional (1D) system describing the re-parameterization invariant FRW model. This investigation focuses on deriving nilpotent and anti-commutative BRST-anti-BRST symmetry transformations,  associated conserved BRST-anti-BRST Noether charges and their nilpotency and absolute properties, and invariance of the Lagrangian.
From a physical standpoint, the nilpotency of these symmetries and charges highlights their fermionic characteristics, while their absolute anti-commutativity underscores the linear independence of these nilpotent symmetries and charges. These features are crucial for understanding the structure and consistency of the BRST formalism in the context of this 1D system.
In this work, we have successfully encapsulated all the aforementioned key features within the framework of the ACSA to BRST formalism. This has been achieved by considering only the  (anti-)chiral supervariables and their corresponding super expansions along the Grassmannian directions of the (1, 1)-dimensional  (anti-)chiral super sub-manifolds embedded within the most general (1, 2)-dimensional supermanifold. 
A notable result of our analysis is the demonstration of the absolute anti-commutativity of the conserved, nilpotent BRST-anti-BRST charges within the ACSA framework, achieved solely using the  (anti-)chiral super expansions of the corresponding supervariables.

\vskip 0.15 cm

The structure of our present work is as follows. Section 2 introduces re-parameterization symmetries and constraints analysis of the Lagrangian that describe the cosmological FRW model. In Section 3, we explore the BRST-anti-BRST symmetries corresponding to scale factor and lapse function and derive the CF-type restriction using MBTSA to BRST formalism.  Section 4 focuses on deriving the BRST-anti-BRST symmetries within the ACSA to BRST formalism, where quantum gauge (i.e., BRST-anti-BRST) invariant restrictions on the  (anti-)chiral supervariables play a pivotal role. Section 5 is dedicated to deriving the final re-parameterization invariant Lagrangian by BRST-anti-BRST symmetries.   In Section 6, we demonstrate the existence of the Curci-Ferrari (CF)-type restriction by examining the symmetry invariance of the charges and showing the nilpotency and absolute anti-commutativity properties within the framework of ACSA. Section 7 captures the Lagrangian invariance
of the FRW model and demonstrates the CF-type restriction by employing key techniques from the ACSA to the BRST formalism. Finally, Section 8 provides concluding remarks 
 on our cosmological theory and suggests some important potential directions for future research. In Appendix A, we discuss the cosmological FRW model with the differential gauge condition.\\

\vskip 0.1 cm

\section{Preliminaries: Lagrangian  and Re-parameterization  Symmetries of the Cosmological FRW Model} 

\vskip 0.1 cm

In this section, we review the fundamentals of the cosmological FRW model which describes a homogeneous and isotropic universe. We start with the FRW metric tensor, expressed in spherical coordinates\footnote{The coordinates \( r \), \( \vartheta \), and \( \phi \) represent comoving radial and angular positions. This means that they remain fixed concerning time for objects that are at rest relative to the expanding space. The physical distances between objects change as the universe expands, but their comoving coordinates do not. The time $t$ is the cosmic time which describes the time elapsed according to an observer who is stationary concerning the expansion of the universe (i.e., an observer who is at rest in comoving coordinates).
} \((t, r, \vartheta, \phi)\):
\begin{equation}
ds^2 = N^2(t) \, dt^2 + a^2(t) \left(\frac{1}{1 - k r^2}\right) dr^2 + a^2(t) \left(r^2 d\vartheta^2 + r^2 \sin^2 \vartheta \, d\phi^2\right),
\end{equation}
where \(N(t)\) is the lapse function, and \(a(t)\) represents the scale factor which characterizes the size and expansion of the universe on large scales. The curvature parameter \(k\) can take values \(0\), \(-1\), or \(+1\), corresponding to a flat (\(k=0\)), open (\(k=-1\)), or closed (\(k=+1\)) universe, respectively.

The classical Lagrangian for this model, expressed in terms of Arnowitt-Deser-Misner (ADM) variables, is given by:
\begin{equation}
L_s = -\frac{1}{2} \frac{a \dot{a}^2}{N} + \frac{1}{2} k N a,
\end{equation}
where \(\dot{a}\) denotes the time derivative of scale factor  \(a(t)\). 
This Lagrangian represents a second-order form. However, the corresponding first-order Lagrangian (\(L_f\)) can be derived using the Legendre transformation and is expressed as: 
\begin{equation}
L_f = p_a \dot{a} + \frac{1}{2a} p_a^2 N + \frac{1}{2} k N a,
\end{equation}
where \(p_a\) (defined as \(p_a = -\frac{a \dot{a}}{N}\)) is the canonical conjugate momentum associated with the scale factor \(a\).
The Euler-Lagrange equations of motion (EL-EoMs) derived from the above Lagrangians (2) and (3) are:
\begin{align}
& \dot{a}^2 + k N^2 = 0, \quad a \ddot{a} + k N^2 + \frac{\dot{a}^2}{2} 
- \frac{a \dot{a}}{N} = 0, \\
& a \, \dot a  + p_a \, N = 0, \quad a^2\, (k\, N + 2\, \dot p_a) + p_a{^2} \, N = 0, \quad 
 p_a{^2} + a^2\, k = 0.
\end{align}
Equations (4) and (5) are the EL-EoMs of Lagrangians $L_s$ and $L_f$, respectively. 
The canonical momentum associated with the lapse function \(N\) is zero because \(N\) does not appear with a time derivative in the Lagrangian:
\begin{equation}
\Pi_{(N)} \approx 0.
\end{equation}
This represents a primary constraint of the system. The canonical Hamiltonian is given by:
\begin{equation}
H = p_a \dot{a} - L_f = -\frac{N p_a^2}{2a} - \frac{1}{2} k N a,
\end{equation}
where \(p_a\) is the momentum conjugate to \(a\).
The time conservation of the primary constraint leads to the secondary constraint:
\begin{equation}
\frac{d \Pi_{(N)}}{dt} = \frac{p_a^2}{2a} + \frac{k}{2} a \approx 0.
\end{equation}
Both the primary and secondary constraints are first-class since they commute. This indicates the presence of gauge symmetry in the FRW model.

The gauge transformations for the variables \(N\), \(a\) and $p_a$ are given by:
\begin{align}
\delta_g N = N \,\dot{\eta} + \dot{N} \,\eta, \quad \delta_g a = \dot{a} \,\eta,\quad 
\delta_g p_a = \dot p_a \, \eta,
\end{align}
where \(\eta(t)\) is an infinitesimal gauge transformation parameter. Under these transformations, the Lagrangian  (3) transforms as a total time derivative:
\begin{equation}
\delta_g L_f = \frac{d}{dt} \left(\eta L_f\right).
\end{equation}
This implies that the Lagrangian remains invariant under these gauge transformations which ensures the invariance of the corresponding action integral.
We end this section with the remarks that symmetry transformations given in (9)
are also infinitesimal re-parameterization symmetry transformations because of the time re-parameterization as $ t \longrightarrow t^\prime  = g(t)$ where $g(t)$ is a physically well-defined function of $t$. However, this function is taken as $g(t)  = t - \epsilon (t)$ for its infinitesimal version where $\epsilon (t)$  is the infinitesimal parameter.\\


\section {BRST-anti-BRST  Symmetries   for Target Space Variables and CF-Type Restriction: MBTSA} 

\vskip 0.1 cm

In this section, we derive the off-shell nilpotent BRST-anti-BRST symmetries for the target space variables \(a(t)\) and \(p_a (t)\) using the modified Bonora-Tonin (BT) supervariable approach (MBTSA) within the BRST framework [36-40]. This method considers super diffeomorphism transformations [see Eq. (11) below] along with the complete expansion of supervariables across all Grassmannian directions of the \((1, 2)\)-dimensional supermanifold.

\vskip 0.15 cm

To derive the BRST-anti-BRST symmetries for the target space phase variables [i.e., \(a(t)\) and \(p_a (t)\)], we first generalize the re-parameterization (i.e., diffeomorphism) symmetry transformation parameter \(t\) [i.e., \(t \longrightarrow t' = g(t) \equiv t - \epsilon(t)\)] from the ordinary 1D spacetime manifold onto our suitably chosen \((1, 2)\)-dimensional supermanifold.
\begin{eqnarray}
g(t) \longrightarrow \tilde g(t, \vartheta, \bar\vartheta)  
=  t - \vartheta\;\bar C (t) - \bar\vartheta\; C(t) +  \vartheta\,\bar\vartheta\,h (t),
\end{eqnarray} 
where the supermanifold is parameterized by the variables \((t, \vartheta, \bar\vartheta)\), and the infinitesimal parameter \(\epsilon(t)\) is replaced by the fermionic (anti-)ghost variables \((\bar{C}, C)\). These variables are incorporated into Eq. (11) as coefficients of \((\vartheta,  \bar{\vartheta})\) due to the connection between the Grassmannian translational generators \((\partial_\vartheta, \partial_{\bar{\vartheta}})\) along the \((\vartheta, \bar{\vartheta})\)-directions and the off-shell nilpotent quantum  BRST-anti-BRST symmetry transformations \(s_{(a)b}\) [17, 18]. Specifically, we have already embedded the BRST-anti-BRST symmetry transformations \(s_{ab} t = -\,\bar{C}\) and \(s_b t = -\,C\) into the expansion in Eq. (11). To fully determine the expression for the derived variable \(h(t)\) present in the Eq. (11), further consistency conditions are required.

\vskip 0.15 cm

According to the basic tenets of the  {modified} BT-  supervariable approach to BRST formalism, all the  {ordinary} variables of the theory must be generalized  {onto} a suitably chosen \((1, 2)\)-dimensional supermanifold  {as} supervariables. In this process, the  {generalization} in Eq. (11) is included as  {one} of the arguments of these supervariables. Following this, we consider the  {full} super-expansions along  {all} possible Grassmannian directions of the \((1, 2)\)-dimensional supermanifold. Consequently, we obtain the following explicit generalizations for the target space variables:
\begin{eqnarray}
a(t) &\longrightarrow & \tilde{A} (\tilde{g}(t, \vartheta, \bar{\vartheta}), \vartheta, \bar{\vartheta})\nonumber\\
 & = & A (\tilde{g}(t, \vartheta, \bar{\vartheta})) + \vartheta\,\bar{R}(\tilde{g}(t, \vartheta, \bar{\vartheta})) +  \bar{\vartheta}\,{R}(\tilde{g}(t, \vartheta, \bar{\vartheta})) + \vartheta\,\bar{\vartheta}\,S(\tilde{g}(t, \vartheta, \bar{\vartheta})),\nonumber\\
p_a (t) & \longrightarrow & \tilde{P}_a(\tilde{g}(t, \vartheta, \bar{\vartheta}), \vartheta, \bar{\vartheta}) \nonumber\\
& = & P_a(\tilde{g}(t, \vartheta, \bar{\vartheta})) + \vartheta\,\bar{T}(\tilde{g}(t, \vartheta, \bar{\vartheta})) +  \bar{\vartheta}\,{T} (\tilde{g}(t, \vartheta, \bar{\vartheta})) + \vartheta\,\bar{\vartheta}\,U (\tilde{g}(t, \vartheta, \bar{\vartheta})),
\end{eqnarray}
where all the derived  supervariables on the right-hand side (i.e., \(R, \bar{R}, S, T, \bar{T}, U \)) are functions of the super diffeomorphism transformation in Eq. (11). Therefore, we need to perform the appropriate Taylor expansion of all these supervariables as follows:
\begin{eqnarray}
&&A (t - \vartheta\,\bar{C} - \bar{\vartheta}\,C + \vartheta\,\bar{\vartheta}\,h)  =  a(t) - \vartheta\,\bar{C}\,\dot{a} 
- \bar{\vartheta}\,C\,\dot{a} +  \vartheta\,\bar{\vartheta}\,(h\,\dot a - \bar{C}\,C\,\ddot{a}), \nonumber \\
&&\vartheta\,\bar{R} (t - \vartheta\,\bar{C} - \bar{\vartheta}\,C + \vartheta\,\bar{\vartheta}\, h) = \vartheta\,\bar{R}(t) - \vartheta\,\bar{\vartheta}\,C\,\dot{\bar{R}} (t), \nonumber \\
&&\bar{\vartheta}\,{R} (t - \vartheta\,\bar{C} - \bar{\vartheta}\,C + \vartheta\,\bar{\vartheta}\,h) = \bar{\vartheta}\,{R} (t) + \vartheta\,\bar{\vartheta}\,\bar{C}\,\dot{{R}} (t), \nonumber \\
&&\vartheta\,\bar{\vartheta}\,S (t - \vartheta\,\bar{C} - \bar{\vartheta}\,C + \vartheta\,\bar{\vartheta}\,h) = \vartheta\,\bar{\vartheta}\,S(t). 
\end{eqnarray}
In the above, we have considered the usual key properties of the Grassmannian variables \((\vartheta, \bar{\vartheta})\), namely \(\vartheta^2 = \bar{\vartheta}^2 = 0\) and \(\vartheta \bar{\vartheta} + \bar{\vartheta} \vartheta = 0\). Similarly, we need to expand the derived supervariables in the expression for \(\tilde{P_a} (\tilde{g}(t, \vartheta, \bar{\vartheta}), \vartheta, \bar{\vartheta})\). This can be expressed as follows:
\begin{eqnarray}
&&P_a(t - \vartheta\,\bar{C} - \bar{\vartheta}\,C + \vartheta\,\bar{\vartheta}\,h) = p_a (t) - \vartheta\,\bar{C}\,\dot{p}_a 
- \bar{\vartheta}\,C\,\dot{p}_a  + \vartheta\,\bar{\vartheta}\,(h\,\dot{p}_a - \bar{C}\,C\,\ddot{p}_a), \nonumber \\
&&\vartheta\,\bar{T}(t - \vartheta\,\bar{C} - \bar{\vartheta}\,C + \vartheta\,\bar{\vartheta}\,h) = \vartheta\,\bar{T}(t) - \vartheta\,\bar{\vartheta}\,C\,\dot{\bar{T}} (t), \nonumber \\
&&\bar{\vartheta}\,{T} (t - \vartheta\,\bar{C} - \bar{\vartheta}\,C + \vartheta\,\bar{\vartheta}\,h) = \bar{\vartheta}\,{T} (t) + \vartheta\,\bar{\vartheta}\,\bar{C}\,\dot{{T}} (t), \nonumber \\
&&\vartheta\,\bar{\vartheta}\,U (t - \vartheta\,\bar{C} - \bar{\vartheta}\,C + \vartheta\,\bar{\vartheta}\,h) = \vartheta\,\bar{\vartheta}\,U (t).
\end{eqnarray}
{ In the super expansion of supervariable, the connection between the BRST-anti-BRST symmetry transformations ($s_b, \, s_{ab}$) and the Grassmannian partial derivatives  $(\partial_{\bar\vartheta}, \, \partial_{\vartheta})$ on (1, 2)-dimensional supermanifold have been established through the mapping: $s_b \longleftrightarrow \partial_{\bar\vartheta}$ and  $s_{ab} \longleftrightarrow \partial_{\vartheta}$  (see, e.g., [17-20] for details). More specifically, the BRST-anti-BRST symmetry transformations of any generic variable $\psi(t)$ correspond to the translation of the associated supervariable  $\Psi(t, \vartheta, \bar\vartheta)$ along the $\bar\vartheta$ and $\vartheta$-directions. This relationship can be mathematically expressed as $
s_b \psi (t) = \frac{\partial}{\partial \bar\vartheta} \Psi(t, \vartheta, \bar\vartheta) = \partial_{\bar\vartheta} \Psi^{(b)}(t, \vartheta, \bar\vartheta)$ at $\vartheta = 0$ and 
$s_{ab} \psi(t) = \frac{\partial}{\partial \vartheta} \Psi(t, \vartheta, \bar\theta) = \partial_{\vartheta} \Psi (t, \vartheta, \bar\theta)$ at $\bar\vartheta = 0$.  In other words, the coefficient of $\bar\vartheta$ in the super expansion of supervariable directly represents the BRST symmetry transformation of the corresponding variable. The  coefficient of $\vartheta$  in the super expansion of supervariable directly represents the anti-BRST symmetry transformation of the corresponding variable.
Moreover, in super expansion the coefficient of $\vartheta\,\bar\vartheta$ represents the $s_b\,s_{ab}$. Thus, the  anti-commutativity of the Grassmannian derivatives $(\partial_{\bar\vartheta}, \partial_\vartheta + \partial_{\vartheta}, \partial_{\bar\vartheta} = 0)$ implies  that the  absolute anti-commutativity of the BRST-anti-BRST symmetries   (i.e. $s_b\, s_{ab}  + s_{ab}\,s_b = 0$)}.

Ultimately, the secondary supervariables on the right-hand side of Eq. (12) must be replaced by the ordinary derived variables, as they are Lorentz scalars with respect to the 1D spacetime manifold (i.e., the 1D trajectory of a cosmological FRW model of universe embedded in a \(D\)-dimensional target flat Minkowskian spacetime manifold). 
Consequently, the final expressions for the super expansions  in Eq. (12) are:
\begin{eqnarray}
\tilde{A}(\tilde{g}(t, \vartheta, \bar{\vartheta}), \vartheta, \bar{\vartheta})  & = & a (t) + \vartheta\,\bar{R} (t) 
+ \bar{\vartheta}\,{R} (t)+  \vartheta\,\bar{\vartheta}\,S (t),\nonumber\\
&\equiv & a (t) + \vartheta\,(s_{ab}\,a  (t))  +  \bar{\vartheta}\,( s_b\,a  (t)) +  \vartheta\,\bar{\vartheta}\,(s_b\,s_{ab}\,a  (t)),\nonumber\\
\tilde{P}_a (\tilde{g}(t, \vartheta, \bar{\vartheta}), \vartheta, \bar{\vartheta})  & = & p_a (t) + \vartheta\,\bar{T} (t) 
+ \bar{\vartheta}\,{T} (t) +  \vartheta\,\bar{\vartheta}\,U (t),\nonumber\\
&\equiv & p_a (t) + \vartheta\,(s_{ab}\,p_a  (t)) + \bar{\vartheta}\,( s_b\,p_a  (t)) +     \vartheta\,\bar{\vartheta}\,(s_b\,s_{ab}\,p_a  (t)).
\end{eqnarray} 
It is clear that we need to explicitly compute the exact values of the derived variables \(\big[R (t), \bar{R} (t), S (t), T (t), \bar{T} (t), U (t)\big]\), present in the expansions of Eq (15), to derive the nilpotent BRST-anti-BRST symmetry transformations [\(s_{(a)b}\)] for the variables \(a (t)\) and \(p_a (t)\).

\vskip 0.15 cm

As a side remark, we emphasize that for the proper off-shell BRST-anti-BRST symmetries to exist, the conditions \(s_b\,s_{ab}\,a (t) = -\,s_{ab}\,s_b\,a (t)\) and \(s_b\,s_{ab}\,p_a (t) = -\,s_{ab}\,s_b\,p_a (t)\) must be satisfied. These conditions ensure the absolute anti-commutativity property  (i.e., \(s_b\,s_{ab} + s_{ab}\,s_b = 0\)) of the off-shell BRST-anti-BRST symmetry transformations \([s_{(a)b}]\).

\vskip 0.15 cm

{Now we exploit the theoretical strength of the ``horizontality condition" HC (in the context of our present set of 1D diffeomorphism invariant theories) which is nothing but the requirement that a
Lorentz scalar does not transform under any kind of internal, spacetime, supersymmetric,
diffeomorphism, etc., transformations.} 
At this point, we  use the theoretical strength   and power of the HC in the re-parameterization invariant theory and impose the following 
requirement on physical grounds
\begin{eqnarray}
&& \tilde{A} (\tilde{g}(t, \vartheta, \bar{\vartheta}), \vartheta, \bar{\vartheta})   \equiv {A}(t, \vartheta, \bar{\vartheta})= a (t), \nonumber\\
&& \tilde{P}_a (\tilde{g}(t, \vartheta, \bar{\vartheta}),\vartheta, \bar{\vartheta})   \equiv {P}_a (t, \vartheta, \bar{\vartheta})= p_a (t),
\end{eqnarray} 
due to the fact that \(a (t)\) and \(p_a (t)\) are  {scalars} with respect to the 1D diffeomorphism transformation. For the above equality to hold, we must systematically and precisely collect  {all} the expansions in Eqs. (13) and (14), as illustrated below:
\begin{eqnarray}
\tilde{A} (\tilde{g}(t, \vartheta, \bar{\vartheta}), \vartheta, \bar{\vartheta})   &=&  a (t) + \vartheta\, (\bar R - \bar C\,\dot a) + \bar\vartheta\, (R - C\,\dot a)\nonumber\\
& + &   \vartheta\,\bar\vartheta\,\big [ S + \bar C\,\dot{R} -  C\,\dot{\bar R} + h\; \dot a  - \bar C\,C\,\ddot a \big],\nonumber\\
 \tilde{P}_a (\tilde{g}(t, \vartheta, \bar{\vartheta}), \vartheta, \bar{\vartheta})   & = & p_a (t) + \vartheta\, (\bar T - \bar C\,\dot p_a)  + \bar\vartheta\, (T - C\,\dot p_a)\nonumber\\
& + &   \vartheta\,\bar\vartheta\,\big [ U + \bar C\,\dot{T} -  C\,\dot{\bar T} + h \; \dot p_a  - \bar C\,C\,\ddot p_a \big].
\end{eqnarray}
Now, we leverage the theoretical potential and power of the HC in the re-parameterization invariant theory. Mathematically, this requires that:
\begin{eqnarray}
\tilde{A} (\tilde{g}(t, \vartheta, \bar{\vartheta}), \vartheta, \bar{\vartheta}) = a (t), \quad \tilde{P}_a  (\tilde{g}(t, \vartheta, \bar{\vartheta}), \vartheta, \bar{\vartheta}) = p_a (t).
\end{eqnarray}
This leads to the determination of the  {secondary} variables as:
\begin{eqnarray}
&&\bar{R} = \bar{C}\,\dot{a},\qquad R = C\,\dot{a}, \qquad  S = C\,\dot{\bar{R}} - \bar{C}\,\dot{R} + \bar{C}\,C \ddot{a} - \; h\, \dot a, \nonumber \\
&&\bar{T} = \bar{C}\,\dot{p}_a, \qquad T = C\,\dot{p}_a,\qquad U = C\,\dot{\bar{T}} - \bar{C}\,\dot{T} + \bar{C}\,C \,\ddot{p}_a
- \, h \, \dot p_a.
\end{eqnarray}
Plugging in the values of $R, \bar{R}, T$ and $\bar{T}$ in the above, we obtain the following expressions for $S(t)$ and $U(t)$, namely:
\begin{eqnarray} 
&&S(t) = -[(\dot{\bar{C}}\,C + \bar{C}\,\dot{C} + h)\,\dot{a} + \bar{C}\,C\,\ddot{a}], \nonumber \\
&&U(t) = -[(\dot{\bar{C}}\,C + \bar{C}\,\dot{C} + h)\,\dot{p}_a + \bar{C}\,C\,\ddot{p}_a].
\end{eqnarray}
As previously discussed, the above expressions are also equivalent to \(s_b\,s_{ab}\,a = -s_{ab}\,s_b\,a\) and \(s_b\,s_{ab}\,p_a = -s_{ab}\,s_b\,p_a\), respectively. We have already derived that \(s_b\,a = C\,\dot{a}\), \(s_{ab}\,a = \bar{C}\,\dot{a}\), \(s_b\,p_a = C\,\dot{p}_a\), and \(s_{ab}\,p_a = \bar{C}\,\dot{p}_a\), based on the comparison with Eq. (15). Therefore, using Eq. (19), the expressions for \(R\), \(\bar{R}\), \(T\), and \(\bar{T}\) indicate that we have already derived the nilpotent BRST-anti-BRST symmetries for the target space variables \(a(t)\) and \(p_a (t)\).

\vskip 0.15 cm

The nilpotency $[s_{(a)b}^2 = 0]$ properties of $s_{(a)b}$ lead to the derivation 
of the BRST-anti-BRST symmetry transformations on the (anti-)ghost variables as:
\begin{eqnarray}
s_b\,C = C\,\dot{C}, \qquad\qquad s_{ab}\,\bar{C} = \bar{C}\,\dot{\bar{C}}.
\end{eqnarray}  
We assume that \(s_{ab}\,C = \bar{\cal B}\) and \(s_b\,\bar{C} = {\cal B}\), where \({\cal B}\) and \(\bar{\cal B}\) are the auxiliary variables of the Nakanishi-Lautrup type of the theory. These transformations (i.e., \(s_b\,\bar{C} = {\cal B}\), \(s_{ab}\,C = \bar{\cal B}\)) are the standard assumptions within the BRST formalism. As a result of these off-shell nilpotent transformations (i.e., \(s_{(a)b}^2 = 0\)), we observe the following:
\begin{eqnarray}
s_b\,s_{ab}\,a & = & ({\cal B}- \bar{C}\,\dot{C})\,\dot a - \bar{C}\,C\,\ddot a \equiv S (t), \nonumber \\
- s_{ab}\,s_b\,a & = &  (- \bar{\cal B} - \dot{\bar{C}}\,{C})\,\dot a - \bar{C}\,C\,\ddot a \equiv S (t), \nonumber \\
s_b\,s_{ab}\,p_a & = & ({\cal B}- \bar{C}\,\dot{C})\,\dot{p}_a - \bar{C}\,C\,\ddot{p}_a \equiv U (t), \nonumber \\
- s_{ab}\,s_b\,p_a & = & (- \bar{\cal B} - \dot{\bar{C}}\,{C})\,\dot{p}_a - \bar{C}\,C\,\ddot{p}_a \equiv U (t).
\end{eqnarray}
In the comparison of the above with the expressions in Eq. (20) (derived from the MBTSA), we derive the derived variable \(h(t)\) as
\begin{eqnarray}  
h(t) & = & \bar{\cal B} - \bar{C}\,\dot{C}  \equiv  - {\cal B} - \dot{\bar{C}}\,C \quad  \Longrightarrow \quad   {\cal B} + \bar{\cal B} + (\dot{\bar{C}}\,C - \bar{C}\,\dot{C}) = 0.
\end{eqnarray}
Thus, we have obtained the celebrated Curci-Ferrari (CF)-type restriction from the application of MBTSA to the BRST formalism. This is the determination of the derived variable \(h (t)\) [cf. Eq. (11)], expressed in terms of the basic and auxiliary variables, derived from the requirement (that is, \(s_b\,s_{ab}\,a = -\,s_{ab} \,s_b\,a\) or \(s_b\,s_{ab}\,p_a = -\,s_{ab} \,s_b\,p_a\)) of absolute anti-commutativity \((s_b\,s_{ab} + s_{ab} \,s_b = 0)\) of the \(s_{(a)b}\) symmetries which play very crucial role.

\vskip 0.15 cm

We conclude this section with the following remarks. First, we observe that our choice of \(s_b\, \bar{C} = {\cal B}\) and \(s_{ab}\,C = \bar{\cal B}\) implies the following generalizations for the (anti-)ghost variables \((\bar{C}) C\) from the 1D  {ordinary} spacetime manifold to the (1, 1)-dimensional  {chiral-anti-chiral } super sub-manifold of the (1, 2)-dimensional supermanifold, as
\begin{eqnarray}  
C (t) \longrightarrow F^{(c)} (t, \vartheta) &=& C(t) + \vartheta\, [{\bar {\cal B}} (t)]\equiv   C (t) + \vartheta\, (s_{ab}C),\nonumber\\
~~~~\bar C (t) \longrightarrow \bar F^{(ac)} (t, \bar\vartheta) & = & \bar C(t) + \bar\vartheta\, [{\cal B} (t)] \equiv   \bar C (t)  + \bar\vartheta\, (s_b\bar C),
\end{eqnarray}
where the superscripts \((c)\) and \((ac)\) denote the  {chiral} and  {anti-chiral} super expansions. This observation subtly suggests that the  (anti-)chiral supervariable approach (ACSA) to the BRST formalism [27-33] will be instrumental in our further discussions. Additionally, it can be verified that the absolute anti-commutativity property \((s_b\, s_{ab} + s_{ab}\, s_b = 0)\), for the phase space target variables \([a (t), p_a (t)]\), holds with respect to the off-shell nilpotent BRST-anti-BRST symmetry transformations, namely:
\begin{eqnarray}
&&\{s_b, s_{ab}\}\, a = [{\cal B} + {\bar {\cal B}} + (\dot{\bar C}\,C - \bar C\,\dot C)]\;\dot a = 0,\nonumber\\
&&\{s_b, s_{ab}\}\, p_a = [{\cal B} + {\bar {\cal B}} + (\dot{\bar C}\,C - \bar C\,\dot C)]\;\dot p_a = 0,
\end{eqnarray}
are valid if and only if we invoke the power and potential of the CF-type restriction (23) from an external perspective. Finally, we emphasize that the condition for the absolute anti-commutativity of the BRST-anti-BRST symmetry transformations \(s_{(a)b}\) with respect to the (anti-)ghost variables, namely:
\begin{eqnarray}
&&\{s_b, s_{ab}\}\,C = 0 \;\Longrightarrow \;s_b {\bar {\cal B}} = \dot{{\bar {\cal B}}}\;C - {\bar {\cal B}}\,\dot C,\nonumber\\
&&\{s_b, s_{ab}\}\,\bar C = 0\;\Longrightarrow \; s_{ab}  B = \dot{\cal B}\;\bar C -  B\,\dot{\bar C},
\end{eqnarray} 
leads to the derivation of \(s_b \bar{\cal B} = \dot{\bar{\cal B}}\,C - \bar{\cal B}\,\dot{C}\) and \( s_{ab} {\cal B} = \dot{\cal B}\,\bar{C} - {\cal B}\,\dot{\bar{C}}\), which are found to be off-shell nilpotent (\([s_{(a)b}]^2 = 0\)) and absolutely anticommuting in nature
(i.e., $(\{s_b, s_{ab}\}\,{\cal B} = 0\), \(\{s_b, s_{ab}\}\,\bar{\cal B} = 0)$)
without any reliance on the CF-type restriction.\\


\section {BRST-anti-BRST Symmetries: ACSA}

\vskip 0.1 cm

As mentioned previously, we have already employed the   (anti-)chiral supervariable approach (ACSA) to determine the BRST-anti-BRST symmetry transformations: \( s_{ab} C = \bar{\cal B} \) and \( s_b \bar{C} = {\cal B} \) [refer to Eq. (24)], which are fundamental assumptions within the BRST formalism.

\vskip 0.15 cm

In this section, we extend the application of ACSA to derive the remaining off-shell nilpotent BRST-anti-BRST symmetry transformations, building on our earlier results. These prior results include \( s_b a = C \dot{a} \), \( s_{ab} a = \bar{C} \dot{a} \), \( s_b p_a = C \dot{p}_a \), \( s_{ab} p_a = \bar{C} \dot{p}_a \), \( s_b \bar{C} = {\cal B} \), and \( s_{ab} C = \bar{\cal B} \).
To accomplish this goal, we extend the one dimensional ordinary variables \([N(t), C(t), \bar{\cal B}(t), {\cal B} (t)]\) onto a \((1, 1)\)-dimensional anti-chiral super sub-manifold of the broader \((1,2)\)-dimensional supermanifold. This generalization is crucial for deriving the complete set of symmetry transformations in a consistent and comprehensive manner.
\begin{eqnarray} 
&&N (t)\; \longrightarrow \tilde N(t, \bar\vartheta) = N (t) + \bar\vartheta\,g_1(t), \nonumber\\
&&{\cal B} (t) \;\longrightarrow \tilde{\cal B}(t, \bar\vartheta) = {\cal B} (t) + \bar\vartheta\,g_2(t), \nonumber\\
&&\bar{\cal B}(t)\; \longrightarrow {\tilde{\bar{\cal B}}}(t, \bar\vartheta) = {\bar {\cal B}}(t) + \bar\vartheta\,g_3(t),\nonumber\\
&& C(t) \longrightarrow F(t, \bar\vartheta) = C(t) + \bar\vartheta\,b_1(t), 
\end{eqnarray}
The derived  variables \((g_1, g_2, g_3)\) are  {fermionic}, while \(b_1\) is  {bosonic}, owing to the fermionic nature of \(\bar{\vartheta}\) (as \(\bar{\vartheta}^2 = 0\)). It is clear that in the limit \(\bar{\vartheta} = 0\), the super-expansions reduce to the  {ordinary} variables \([N(t), C(t), {\cal B} (t), \bar{\cal B}(t)]\). 
Furthermore, our \((1,1)\)-dimensional  {anti-chiral} super sub-manifold is parameterized by \((t, \bar{\vartheta})\), where \(t\) is a  {bosonic} evolution parameter.

\vskip 0.15 cm

To determine the  {secondary} variables in terms of the  {basic} and  {auxiliary} variables of the theory, we utilize one of the fundamental principles of the ACSA, which asserts that  {quantum} gauge (i.e., BRST) invariant quantities must be independent of the Grassmannian variable \(\bar{\vartheta}\). In this context, we observe:
\begin{eqnarray}
&& s_b\,(C\,\dot{a}) = 0, \quad s_b\,(N\,\dot{C} + \dot{N}\,C) = 0, \quad  s_b\,{(\dot{\bar{\cal B}}}\,C - \bar{\cal B}\,\dot{C}) = 0, \quad s_b\,{\cal B} = 0.
\end{eqnarray} 
The BRST invariant quantities, when extended onto a \((1,1)\)-dimensional  {anti-chiral} super sub-manifold, must remain independent of \(\bar{\vartheta}\). In other words, the following equalities hold:
\begin{eqnarray}
&&F(t, \bar{\vartheta})\,\dot{A}^{(ha)}(t, \bar{\vartheta}) = C(t)\,\dot{a}(t), \quad\tilde{\cal B}(t, \bar{\vartheta}) = {\cal B} (t), \nonumber\\
&& \tilde N(t, {\bar{\vartheta}})\,\dot{F}(t, \bar{\vartheta})  + \dot{\tilde N}(t, \bar{\vartheta})\,F(t, \bar{\vartheta}) 
= N(t)\,\dot{C}(t) + \dot{N}(t)\,C(t),\nonumber\\
&& {\dot{\tilde{\bar{\cal B}}}}(t, \bar{\vartheta}) \, F(t, \bar{\vartheta}) - {\tilde{\bar{\cal B}}}(t, \bar{\vartheta})\,\dot F(t, \bar{\vartheta})
 = {\dot{\bar{\cal B}}}(t)\,C(t) - \bar{\cal B}(t)\,\dot{C}(t), 
\end{eqnarray}
where $A ^{(ha)}(t, \bar{\vartheta})$ is the {anti-chiral} limit of the {full} super expansion that has been obtained in the { previous} section, namely;
\begin{eqnarray}
A^{(h)}(t, \vartheta, \bar{\vartheta}) & = &  a(t) + {\vartheta}\,(\bar C\,\dot{a}) + \bar\vartheta \,(C\, \dot a) 
 +  \vartheta\,\bar\vartheta\,\big[-\{({\bar {\cal B}} + \dot {\bar C}\,C)\,\dot a + \bar C\,C\, \ddot a\}\big]\nonumber\\
&\equiv & a (t) + {\vartheta}\,(\bar C\,\dot{a} ) + \bar\vartheta \,(C\, \dot a)
+   \vartheta\,\bar\vartheta\;\big[({\cal B}- \bar C\,\dot C)\,\dot a  - \bar C\,C\, \ddot a \big].
\end{eqnarray}
In the above, the superscript \((h)\) indicates that the supervariable \(A^{(h)} (t, \vartheta, \bar{\vartheta})\) has been derived through the application of the HC. This implies the following limiting case (i.e., anti-chiral super expansion):
\begin{eqnarray}
A ^{(ha)}(t, \bar{\vartheta}) & = & a(t)  + \bar\vartheta \,[C (t)\, \dot a(t)],
\end{eqnarray}
The superscript \((ha)\) denotes the  {anti-chiral} limit of the super-expansion (30), which was obtained after applying the HC in the previous section. Substituting from (31) and (27) into the  {first} entry of Eq. (29) leads to \( b_1 (t) = C (t) \dot{C} (t) \). The BRST invariance of the Nakanishi-Lautrup auxiliary variable (i.e., \(  s_b {\cal B} = 0 \)) implies that \( g_2 (t) = 0 \). Therefore, we have the following super-expansions:
\begin{eqnarray} 
&&F ^{(b)}(t, \bar\vartheta)  = C(t) + \bar\vartheta\,(C\,\dot C) \;\equiv \; C(t) + \bar\vartheta\,(s_b\,C(t)), \nonumber\\
&&\tilde{\cal B} ^{(b)}(t, \bar\vartheta) = {\cal B} (t) + \bar\vartheta\,(0)\; \equiv \; {\cal B} (t) + \bar\vartheta\,(s_b\,{\cal B} (t)).
\end{eqnarray}
A closer examination of the above equation reveals that we have already obtained the BRST symmetry transformations:
\( s_b C = C \dot{C} \) and \(  s_b {\cal B} = 0 \), which appear as the coefficients of \(\bar{\vartheta}\) in the expansions (32). The superscript \((b)\) denotes the supervariables obtained after applying the BRST-invariant (i.e.,  {quantum} gauge invariant) restrictions (28). It is worth noting that \( s_b C = C \dot{C} \) can also be derived from the restriction corresponding to the invariance \( s_b (C \dot{p}_a) = 0 \) on the \((1,1)\)-dimensional anti-chiral super sub-manifold. However, for the sake of brevity, we have not discussed this derivation here. In the remaining restrictions of (29), we use the final expressions from (32) to derive the exact expressions for the derived variables as follows:
\begin{eqnarray} 
&&g_1 (t) = N (t)\,\dot C(t) + \dot N (t)\,C(t), \nonumber\\
&& g_3 (t) = {\dot {{\bar {\cal B}}}} (t)\, C(t) - {\bar {\cal B}} (t)\,\dot C (t).
\end{eqnarray}
As a consequence of the above relation (33), we have the following super expansions for { some} of the supervariables [cf. Eq. (27)],  namely; 
\begin{eqnarray} 
&&\tilde N ^{(b)}(t, \bar\vartheta)  = N(t) + \bar\vartheta\,(\dot N\,C + N\,\dot C) \;\equiv \; N (t) + \bar\vartheta\,(s_b\,N (t)), \nonumber\\
&&{\tilde{{\bar {\cal B}}}} ^{(b)}(t, \bar\vartheta) = {\bar {\cal B}}(t) + \bar\vartheta\,{(\dot {{\bar {\cal B}}}}\, C - {\bar {\cal B}}\,\dot C)\; \equiv \; {\bar {\cal B}} (t) + \;\bar\vartheta\,(s_b\,{\bar {\cal B}} (t)),
\end{eqnarray}
The superscript \((b)\) denotes the expansions obtained after applying the BRST (i.e.,  {quantum} gauge) invariance conditions listed in (28). It is straightforward to observe that we have  derived the BRST symmetry  transformations:
\(  s_b {\cal B} = 0 \), \( s_b C = C \dot{C} \), \( s_b N = N \dot{C} + \dot{N} C \), and \( s_b \bar{\cal B} = \dot{\bar{\cal B}} C - \bar{\cal B} \dot{C} \), as the coefficients of the \(\bar{\vartheta}\)-Grassmannian variable in the anti-chiral super-expansions of equations (32) and (34) are the BRST symmetry transformations.

\vskip 0.15 cm

In other words, we see that  {all} the BRST symmetry transformations \((s_b)\) for  {all} variables in our theory is derived in equations (24), (32), and (34),  {except} for the target space variables $a(t)$ and $p_a (t)$ which were obtained earlier by leveraging the theoretical framework of MBTSA to BRST formalism (cf. Sec. 3 for details).

\vskip 0.15 cm

To derive the anti-BRST symmetry transformations \((s_{ab})\) for the variables \(({\cal B}, N, \bar{C}, \bar{\cal B})\), we observe that the following quantities  are anti-BRST invariant, namely:
\begin{eqnarray}
&&s_{ab}\,\bar{\cal B} = 0, \;\; s_{ab}\,(\dot{\cal B}\,\bar{C} - {\cal B}\,\dot{\bar{C}}) = 0, \quad s_{ab}\,(N\,\dot{\bar{C}}  + \dot{N}\,\bar{C}) = 0, \;\; s_{ab}\,(\bar{C}\,\dot{a}) = 0.
\end{eqnarray} 
According to the fundamental principles of ACSA, the above quantities  {must} be independent of the Grassmannian variable \((\vartheta)\) when they are generalized onto a \((1,1)\)-dimensional  {chiral} super sub-manifold of the \((1,2)\)-dimensional supermanifold on which our theory is extended. To achieve this, we generalize the 1D variables \((N, B, \bar{\cal B}, \bar{C})\) onto the selected \((1,1)\)-dimensional  {chiral} super sub-manifold, resulting in the following super-expansions:
\begin{eqnarray}
&& N(t) \longrightarrow \tilde N(t, \vartheta) = N(t) + \vartheta\,\bar{g}_1(t), \nonumber\\
&& {\cal B} (t) \longrightarrow \tilde{\cal B}(t, \vartheta) = {\cal B} (t) + \vartheta\,\bar{g}_2(t), \nonumber \\
&& \bar{\cal B} (t) \longrightarrow {\tilde{\bar{\cal B}}}(t, \vartheta) = \bar{\cal B}(t ) + \vartheta\,\bar{g}_3(t), \nonumber\\
&& \bar{C}(t) \longrightarrow \bar{F}(t, \vartheta) = \bar{C}(t) + \vartheta\,\bar{b}_1(t), 
 \end{eqnarray}
where $\bar{g}_1, \bar{g}_2, \bar{g}_3$ are  {fermionic} and ${\bar{b}}_1$ is the  {bosonic} derived variable due to the fermionic nature of the Grassmannian variable \(\vartheta\) (with \(\vartheta^2 = 0\)), which characterizes the \((1,1)\)-dimensional  {chiral} super sub-manifold. This is in addition to the evolution bosonic parameter \(t\) of our 1D diffeomorphism invariant system. On the other hand, by using the anti-BRST invariant restrictions (35)  and chiral super expansions (36) of the chiral supervariables, we derive the anti-BRST symmetry transformation of our theory which is as follows;
\begin{eqnarray}
s_{ab} N = N\, \dot {\bar C} + \dot N \, \bar C, \quad s_{ab} \bar C = \bar C\, \dot {\bar C}, \quad  s_{ab} {\cal B} = \dot B\, \bar C - {\cal B}\, \dot {\bar C}, \quad s_{ab} {\bar {\cal B}} = 0,
 \end{eqnarray}
Thus, we have derived  BRST-anti-BRST symmetry transformations of all the variables present in our theory using the combination of MBTSA and ACSA to BRST formalism. \\


\section{Lagrangian Formulation: Re-parameterization and Corresponding BRST-anti-BRST Symmetries }

\vskip 0.2 cm

 In our earlier work [30], the complete extended Lagrangian formulation for the cosmological FRW model has been discussed, in detail, with differential gauge conditions (see Appendix A for details).  In this section, we extend the {classical} re-parameterization symmetry \(t \rightarrow {t}' = t - \epsilon(t)\) to its {quantum} counterpart within the framework of BRST formalism. The nilpotent BRST-anti-BRST symmetries, derived in the previous section, play a crucial role in determining the gauge-fixing and Faddeev-Popov (FP) ghost terms in the following manner:
\begin{eqnarray}
s_b\,s_{ab}\Big[\frac{N^2}{2} - \frac{\bar{C}\,C}{2}\Big] = - \frac{{\cal B}^2}{2} -{\cal B} \,\big[N\,\dot{N} + 2\,\dot{\bar{C}}\,C + \bar{C}\,\dot{C}\big]  - N^2\dot{\bar{C}}\,\dot{C} - N\,\dot{N}\,\dot{\bar{C}}\,C - \dot{\bar{C}}\,\bar{C}\,\dot{C}\,C, \nonumber\\
-s_{ab}\,s_b\Big[\frac{N^2}{2} - \frac{\bar{C}\,C}{2}\Big] = - \frac{\bar{\cal B}^2}{2}  + \bar{\cal B}\,\big[N\,\dot{N} + 2\,{\bar{C}}\,\dot{C} + \dot{\bar{C}}\,{C}\big]
  -  N^2\,\dot{\bar{C}}\,\dot{C} - N\,\dot{N}\,{\bar{C}}\,\dot{C} - \dot{\bar{C}}\,\bar{C}\,\dot{C}\,C.
\end{eqnarray} 
As a consequence of \((38)\), we obtain the following BRST-anti-BRST invariant coupled (but equivalent) Lagrangians for our theory:
\begin{eqnarray}
L_{{\cal B}} & = & p_a \dot{a} + \frac{1}{2a} p_a^2 N + \frac{1}{2} k N a
-{\cal B}\;\big(N\,\dot{N} + 2\,\dot{\bar{C}}\,C + \bar{C}\,\dot{C}\big)\nonumber\\
& - & \frac{{\cal B}^2}{2} - N^2\,\dot{\bar{C}}\dot{C} -  N\,\dot{N}\,\dot{\bar{C}}\,C - \dot{\bar{C}}\,\bar{C}\,\dot{C}\,C,\nonumber\\
 L_{\bar{\cal B}} & = & p_a \dot{a} + \frac{1}{2a} p_a^2 N + \frac{1}{2} k N a + \bar{\cal B}\,\big(N\,\dot{N} + 2\,{\bar{C}}\,\dot{C}
 + \dot{\bar{C}}\,{C}\big)\nonumber\\
 & - & \frac{\bar{\cal B}^2}{2} - N^2\,\dot{\bar{C}}\,\dot{C} -  N\,\dot{N}\,{\bar{C}}\,\dot{C} - \dot{\bar{C}}\,\bar{C}\,\dot{C}\,C.
\end{eqnarray}  
We emphasize that the pure FP-ghost component, i.e., $- \dot{\bar{C}} \, \bar{C} \, \dot{C} \, C$, in the Lagrangians (39) remains unchanged. Moreover, due to the off-shell nilpotency of the BRST-anti-BRST symmetries, $s_{(a)b}^2 = 0$, it is evident that $L_{\bar{\cal B}}$ is invariant under the anti-BRST symmetry, and $L_{\cal B}$ is invariant under the BRST symmetry [see Eq. (38)]. To support this statement further, we observe the following:
\begin{eqnarray}
 s_b\, L_{\cal B}  & = &  \frac {d}{dt}\Big [ C\,L_f - N^2\,{\cal B}\,\dot C - N\,\dot N\,{\cal B}\, C  -   {\cal B}\,\bar C\,\dot C\, C -   {\cal B}^2\,C\Big],\nonumber\\
 s_{ab}\, L_{{\bar {\cal B}}} & = & \frac {d}{dt}\Big [ \bar C\,L_f + N^2\,{\bar {\cal B}}\,\dot {\bar C} + N\,\dot N\,{\bar {\cal B}}\, \bar C  
 -   {\bar {\cal B}}\, \dot {\bar C}\,\bar C\,C -  {{\bar {\cal B}}}^2\,\bar C\Big].
\end{eqnarray}
This results in the action integrals $S_1 = \int_{-\infty }^{+\infty } dt \,L_{\bar{\cal B}}$ and $S_2 = \int_{-\infty }^{+\infty } dt \,L_{\cal B}$, which are BRST-anti-BRST invariant, respectively, for the physically well-defined variables that vanish as $t \to \pm \infty$. In the above, the first-order Lagrangian $L_f$ is identical to the one defined in Sec. 2, and the full nilpotent BRST-anti-BRST transformations (for our 1D theory of a cosmological model of the universe) are given by:
\begin{eqnarray}
&&s_b a = C\, \dot a, \;\;  s_b p_a = C\, \dot p_a, \;\; s_b \bar C = {\cal B},\;\; s_b C = C\,\dot C,  \nonumber\\
 && s_b N = \frac {d}{dt}\,(C\,N),\quad  s_b {\cal B} = 0, \quad s_b {\bar {\cal B}} = \dot{{\bar {\cal B}}}\,C - {\bar {\cal B}}\,\dot C, \nonumber\\ 
&&s_{ab} a = \bar C\, \dot a, \,\; s_{ab} p_a = \bar C\, \dot p_a, \,\;\; s_{ab} \bar C = \bar C\,\dot {\bar C},\;\;  s_{ab} C = {\bar {\cal B}},\nonumber\\
&& s_{ab} N = \frac {d}{dt}\,(\bar C\,N),\;\; s_{ab} {\bar {\cal B}} = 0,\quad \;  s_{ab} {\cal B} = \dot{\cal B}\,\bar C - {\cal B}\,\dot {\bar C}.
\end{eqnarray}
The above transformations are off-shell nilpotent, i.e., $s_{(a)b}^2 = 0$, and exhibit absolute anticommutativity. The absolute anti-commutativity property, given by $s_b\, s_{ab} + s_{ab}\, s_b = \{s_b, s_{ab}\} = 0$, holds for all variables in our theory, namely:
\begin{eqnarray}
&& (s_b\, s_{ab}  + s_{ab}\, s_b) \, a = [{\cal B} + {\bar {\cal B}} + (\dot {\bar C}\, C - \bar C\, \dot C)]\;\dot a = 0,\nonumber\\
&&(s_b\, s_{ab}  + s_{ab}\, s_b) \, p_a = [{\cal B} + {\bar {\cal B}} + (\dot {\bar C}\, C - \bar C\, \dot C)]\;\dot p_a = 0,\nonumber\\
&&(s_b\, s_{ab}  + s_{ab}\, s_b) \, N = \frac {d}{dt}\Big[\big\{{\cal B} + {\bar {\cal B}} + (\dot {\bar C}\, C - \bar C\, \dot C)\big\}\,N\Big] = 0,\nonumber\\
&&(s_b\, s_{ab}  + s_{ab}\, s_b) \, C = 0,\quad(s_b\, s_{ab}  + s_{ab}\, s_b) \, \bar C = 0,\nonumber\\
&&(s_b\, s_{ab}  + s_{ab}\, s_b) \, {\cal B} = 0, \;\quad (s_b\, s_{ab}  + s_{ab}\, s_b) \, {\bar {\cal B}} = 0,
\end{eqnarray}
This condition holds provided we impose the BRST-anti-BRST invariant CF-type restriction: 
${\cal B} + \bar{\cal B} + \dot{\bar{C}}\, C - \bar{C}\, \dot{C} = 0$ from outside on our theory. This is, of course, a physical requirement, as it remains invariant under the BRST-anti-BRST symmetry transformations (41). In other words, it can be verified that 
$s_{(a)b} \left[{\cal B} + \bar{\cal B} + (\dot{\bar{C}}\, C - \bar{C}\, \dot{C})\right] = 0$.

\vskip 0.15 cm

As previously stated, the equivalence of the coupled Lagrangians $L_{\cal B}$ and $L_{\bar{\cal B}}$ concerning the off-shell nilpotent BRST-anti-BRST symmetries can be confirmed by the following explicit observations. These arise when we apply $s_b$ to $L_{\bar{\cal B}}$ and $s_{ab}$ to $L_{\cal B}$, namely:
\begin{eqnarray}
s_b\, L_{{\bar {\cal B}}}  & = & \frac {d}{dt}\Big [ C\,L_f + N^2(\dot {\bar C}\, C\, \dot C + {\bar {\cal B}} \,  \dot C) - {\bar {\cal B}} ^2\, C - (2\,{\bar {\cal B}} - {\cal B})\, \bar C\, C\, \dot C +  N\, \dot N \,(\bar C\, C\, \dot C + {\bar {\cal B}} \, C)    \Big]\nonumber\\
& + & ({\cal B}+ {\bar {\cal B}} + \dot {\bar C}\, C - \bar C\, \dot C)\Big[2\,{\bar {\cal B}}\,\dot C - N\,\dot N \, \dot C - 2\,\dot {\bar C}\,C\,\dot C +  \bar C\, \ddot C \, C  \Big]\nonumber\\
& + &  \frac{d}{dt}\Big [{\cal B} +  {{\bar {\cal B}}} + \dot {\bar C}\, C
 - \bar C\, \dot C\Big] \times ({\bar {\cal B}}\, C - N^2\, \dot C), \nonumber\\
 s_{ab}\, L_{\cal B}  & = & \frac {d}{dt}\Big [ \bar C\,L_f + N^2(\dot{\bar C}\,{\bar C}\, \dot C  - {\cal B} \,  \dot {\bar C}) - {\cal B} ^2\, \bar C + (2\,{\cal B} - \bar{\cal B})\dot {\bar C}\, \bar C\, C +  N\,\dot N \,(\dot {\bar C} \,\bar C\, C - {\cal B} \, \bar C)     
 \Big]\nonumber\\
& + & ({\cal B}+ {\bar {\cal B}} + \dot {\bar C}\, C - \bar C\, \dot C)\Big[2\,{\cal B}\,\dot {\bar C} + N\,\dot N \, \dot {\bar C} + 2\,\dot {\bar C}\,\bar C\,\dot C + \ddot{\bar C}\, \bar C \, C \Big]\nonumber\\
& + & \frac{d}{dt}\Big [{\cal B} +  {{\bar {\cal B}}} + \dot {\bar C}\, C - \bar C\, \dot C\Big]\times  ({\cal B}\, \bar C + N^2\, \dot {\bar C}).
\end{eqnarray}
In other words, we observe that both Lagrangians respect both the nilpotent BRST-anti-BRST symmetries [cf. Eq. (41)] provided we account for the validity of the CF-type restriction: 
${\cal B} + \bar{\cal B} + (\dot{\bar{C}}\,C - \bar{C}\,\dot{C}) = 0$. Thus, it is clear that the absolute anti-commutativity property, as well as the equivalence of the Lagrangians $L_{\cal B}$ and $L_{\bar{\cal B}}$, hold true if and only if the CF-type restriction is satisfied. Furthermore, under the validity of the latter, we can express the symmetry transformations as follows:
\begin{eqnarray}
s_{ab}\,L_{\cal B} &=& \frac{d}{dt}\Big[\bar{C}\,L_f - {\cal B}^2\,\bar{C} + N^2(\dot{\bar{C}}\,\bar{C}\,\dot{C} - {\cal B}\,\dot{\bar{C}}) \nonumber\\
& + &  (2\,{\cal B} - \bar{\cal B})\dot{\bar{C}}\,\bar{C}\,C + N\,\dot{N}\,(\dot{\bar{C}}\,\bar{C}\,C - {\cal B}\,\bar{C})  \Big], \nonumber\\
s_b\,L_{\bar{\cal B}} &=& \frac {d}{dt}\Big [ C\,L_f - {\bar {\cal B}}^2\, C + N^2(\dot {\bar C}\, C\, \dot C + {\bar {\cal B}} \,  \dot C), \nonumber\\
& - &  (2\,{\bar {\cal B}} - {\cal B})\, \bar C\, C\, \dot C  + N\,\dot N\,(\bar C\, C\, \dot C + {\bar {\cal B}} \, C) \Big],
\end{eqnarray}
This results in the action integrals $S_1 = \int dt\,L_{\cal B}$ and $S_2 = \int dt\,L_{\bar{\cal B}}$ being BRST-anti-BRST invariant for the physically well-defined variables that vanish as $t \to \pm \infty$, provided our theory is constrained by the BRST-anti-BRST invariant CF-type restriction:
${\cal B} + \bar{\cal B} + \dot{\bar{C}}\,C - \bar{C}\,\dot{C} = 0$.

\vskip 0.15 cm

According to the fundamental principles of Noether's theorem, the above continuous symmetries [i.e., the BRST-anti-BRST symmetries] lead to the derivation of conserved and nilpotent BRST-anti-BRST charges. The equivalent expressions for the conserved BRST charge are:
\begin{eqnarray}
{\cal Q}_b ^{(1)} & = &  N^2\,(\dot {\cal B}\,C - {\cal B}\,\dot C  + \dot {\bar C}\,C\,\dot C) +   2\,N\,\dot N\,\bar C\,C\,\dot C + N^2\,\bar C\, C\, \ddot C,\nonumber\\
&\equiv & s_b [N^2\,(\dot {\bar C}\,C - \bar C\, \dot C)],\nonumber\\
{\cal Q}_b ^{(2)} & = & N^2\,(\bar{\cal B}\,\dot C  - {\dot{\bar{\cal B}}}\,C + 2\,N\,\dot{N}\,\bar{C}\,C\,\dot{C} + 2\,\dot {\bar C}\,C\,\dot C),\nonumber\\
&\equiv & s_{ab}(N^2\,C\,\dot{C}).
\end{eqnarray}
Similarly, the { equivalent} forms of the conserved anti-BRST charge are:
\begin{eqnarray}
{\cal Q}_{ab} ^{(1)} & = &  N^2\,(\bar{\cal B}\,{\dot{\bar{C}}} - {\dot{\bar{\cal B}}}\,\bar{C}+ \dot {\bar C}\,\bar{C}\,\dot C) -   2\,N\,\dot N\,\bar{C}\,\dot{\bar{C}}\,C - N^2\, \bar{C}\, \ddot{\bar{C}}\,C,\nonumber\\
&\equiv & s_{ab} [N^2\,(\bar{C}\,\dot {C} - \dot{\bar{C}}\,{C})], \nonumber\\
{\cal Q}_{ab} ^{(2)} & = & N^2\,(\dot{\cal B}\,\bar{C} - {\cal B}\,\dot{\bar{C}} +2\,N\,\dot{N}\,\dot{\bar{C}}\,\bar{C}\,C + 2\,\dot {\bar{C}}\,\bar{C}\,\dot{C}),\nonumber\\
&\equiv & s_b(N^2\,\dot{\bar{C}}\,\bar{C}).
\end{eqnarray}
The conservation law [i.e., $\dot{{\cal Q}}_{(a)b}^{(r)} = 0$ for $r = 1, 2$] can be proven straightforwardly using the equations of motion derived from the coupled but equivalent Lagrangians  $L_{\cal B}$ and $L_{\bar{\cal B}}$ (39).
We express the conserved charges in BRST-anti-BRST exact forms [cf. Eqs. (45), (46)]. A closer examination of ${\cal Q}_{(a)b}^{(1)}$ reveals the nilpotency of the charges, as it can be seen that:
\begin{eqnarray}
s_b\,{\cal Q}_b^{(1)} = -i\,\{{\cal Q}_b^{(1)}, {\cal Q}_b^{(1)}\} = 0 & \Longrightarrow & ({\cal Q}_b^{(1)})^2 = 0  \Longleftrightarrow   s_b^{2} = 0, \nonumber\\
s_{ab}\,{\cal Q}_{ab}^{(1)} = -i\,\{{\cal Q}_{ab}^{(1)}, {\cal Q}_{ab}^{(1)}\} = 0 & \Longrightarrow & ({\cal Q}_{ab}^{(1)})^2 = 0 \Longleftrightarrow   s_{ab}^{2} = 0.
\end{eqnarray} 
Thus, it is clear from the above equation that the nilpotency of the BRST-anti-BRST symmetries is closely linked to the off-shell nilpotency of the BRST-anti-BRST charges.  Furthermore, we note that the expressions for the conserved charges ${\cal Q}_{(a)b}^{(2)}$ are derived from ${\cal Q}_{(a)b}^{(1)}$ using the elegant and powerful CF-type restriction: 
${\cal B} + \bar{\cal B} + \dot{\bar{C}}\,C - \bar{C}\,\dot{C} = 0$. These expressions for the conserved BRST-anti-BRST charges ${\cal Q}_{(a)b}^{(2)}$ are particularly interesting because they encapsulate the absolute anti-commutativity property.
\begin{eqnarray}
s_{ab}\,{\cal Q}_{b}^{(2)} &=& -i\,\{{\cal Q}_{b}^{(2)}, {\cal Q}_{ab}^{(2)}\} = 0 \quad \Longleftrightarrow  \quad s_{ab}^{2} = 0, \nonumber\\
s_b\,{\cal Q}_{ab}^{(2)} &=& -i\,\{{\cal Q}_{ab}^{(2)}, {\cal Q}_b^{(2)}\} = 0 \quad \Longleftrightarrow  \quad s_b^{2} = 0, 
\end{eqnarray}
We have utilized the fundamental principle connecting the continuous symmetry transformations $s_{(a)b}$ with their generators, which are the conserved Noether BRST-anti-BRST charges. It is important to emphasize that the power and potential of the CF-type restriction have made it possible to express the conserved BRST charge, ${\cal Q}_b^{(2)}$, as an anti-BRST exact quantity, and the conserved anti-BRST charge, ${\cal Q}_{ab}^{(2)}$, as a BRST exact object.

We further discuss the absolute anti-commutativity of the nilpotent BRST-anti-BRST conserved charges and the existence of the BRST-anti-BRST invariant CF-type restriction in our theory. To achieve  the above goal in an alternative manner, we directly apply the BRST symmetry  transformations $(s_b)$ on the expression for the conserved anti-BRST charge 
(${\cal Q}_{ab}^{(1)}$)  as follows:
\begin{eqnarray}
 s_b {\cal Q}_{ab}^{(1)} & = & N^2\, \Big[\frac {d}{d\tau}\Big\{\Big (\frac {d}{d\tau} \big({\cal B}+ {\bar {\cal B}} 
+ \dot{\bar C}\, C - {\bar C}\, \dot C \big)\Big)\, \bar C\, C +   \, {\bar {\cal B}} \,\big({\cal B}+ {\bar {\cal B}}  + \dot{\bar C}\, C - {\bar C}\, \dot C \big)\nonumber\\
& - &  2\, {\dot{\bar {\cal B}}}\, \big({\cal B} + {\bar {\cal B}} +  \dot{\bar C}\, C - {\bar C}\, \dot C \big) -  \big({\cal B}+ {\bar {\cal B}} + \dot{\bar C}\, C -  {\bar C}\, \dot C \big)\dot{\bar C}\, C \Big\} \Big]\nonumber\\
 & - &     \Big\{\frac {d}{d\tau}\Big({\cal B}+ {\bar {\cal B}} +  \dot{\bar C}\, C - {\bar C}\, \dot C \Big)\Big\}\, \bar C  + 2\,N\,\dot N\, C\, \Big[\big({\cal B}+ {\bar {\cal B}} +  \dot{\bar C}\, C - {\bar C}\, \dot C \big)\, \dot {\bar C} \Big].
\end{eqnarray}
 It is evident, from the above, that {every} term on the r.h.s. is { zero} provided we impose the BRST-anti-BRST
 invariant CF-type restriction: ${\cal B} + {\bar {\cal B}} +  \dot{\bar C}\, C - {\bar C}\, \dot C = 0$ from {outside}. In other words, 
 the absolute anti-commutativity of the BRST-anti-BRST charges [hidden in the expression $s_b Q_{ab}^{(1)} \equiv -\,i\,\{{\cal Q}_{ab}^{(1)}, \, {\cal Q}_b^{(1)}\} = 0$ on
the l.h.s. of (48)] is true only in the space of variables where the CF-type restriction of our theory  (${\cal B} + {\bar {\cal B}} +  \dot{\bar C}\, C - {\bar C}\, \dot C = 0$) is satisfied.

\vskip 0.15 cm

To corroborate the above statement, we now apply the anti-BRST symmetry transformation $(s_{ab})$ on the BRST charge ${\cal Q}_{b}^{(1)}$ to obtain the following 
\begin{eqnarray}
 s_{ab} {\cal Q}_{b}^{(1)} & = &  N^2\, \Big[\frac {d}{d\tau}\Big\{\Big (\frac {d}{d\tau} \big({\cal B}+ {\bar {\cal B}} 
+ \dot{\bar C}\, C - {\bar C}\, \dot C \big)\Big)\, \bar C\, C -   {\cal B} \,\big({\cal B}+ {\bar {\cal B}}  + \dot{\bar C}\, C - {\bar C}\, \dot C \big)\nonumber\\
& + &   2\, \dot{\cal B}\, \big({\cal B}+ {\bar {\cal B}} +  \dot{\bar C}\, C - {\bar C}\, \dot C \big) - \,\big({\cal B}+ {\bar {\cal B}} + \dot{\bar C}\, C -  {\bar C}\, \dot C \big){\bar C}\, \dot C \Big\} \Big] \nonumber\\
& - &   \Big\{\frac {d}{d\tau}\Big({\cal B}+ {\bar {\cal B}} +  \dot{\bar C}\, C - {\bar C}\, \dot C \Big)\Big\}\, C -  2\,N\,\dot N\, \bar C\, \Big[\big({\cal B}+ {\bar {\cal B}} +  \dot{\bar C}\, C - {\bar C}\, \dot C \big)\, \dot {C}  \Big],
\end{eqnarray}
which  demonstrates clearly that the absolute anti-commutativity of the BRST-anti-BRST  charges 
is true {only} when the entire  theory is considered on a sub-manifold in the space of {quantum} variables where the BRST-anti-BRST invariant CF-type restriction is satisfied.

\vskip 0.15 cm

In a sense, the above exercise reflects our observations in Eq. (42), where we demonstrated that the absolute anti-commutativity property, $(s_b\,s_{ab} + s_{ab}\,s_b = 0)$, of the BRST-anti-BRST symmetries $s_{(a)b}$ holds true  {only} on a sub-manifold in the space of quantum variables, which is defined by the CF-type equation: 
${\cal B} + \bar{\cal B} + \dot{\bar{C}}\,C - \bar{C}\,\dot{C} = 0$. Since the nilpotency and absolute anti-commutativity properties are essential in the BRST formalism, the requirement of the latter property for the conserved charges in our present discussion leads to the derivation of the CF-type restriction (23), which was also derived from the modified BT-  supervariable approach (MBTSA) to BRST formalism (cf. Sec. 3).

\vskip 0.15 cm

In other words, we directly use the CF-type restriction to recast the conserved BRST-anti-BRST charges in a specific form (e.g., ${\cal Q}_{(a)b}^{(2)}$), such that the BRST charge is expressed as an anti-BRST exact quantity (and the anti-BRST charge as the BRST exact form). At this point, it is clear that the absolute anti-commutativity of $(i)$ the nilpotent BRST-anti-BRST symmetries [cf. Eq. (42)], and $(ii)$ the conserved and nilpotent BRST-anti-BRST charges [cf. Eq. (48)], both originate from the CF-type restriction:
${\cal B} + \bar{\cal B} + \dot{\bar{C}}\,C - \bar{C}\,\dot{C}$ in our 1D diffeomorphism (re-parameterization) invariant cosmological FRW model.\\


\section{Nilpotency and Anti-commutativity Properties of the Conserved BRST-anti-BRST Charges: ACSA}

\vskip .2 cm 

In this section, we capture the off-shell nilpotency as well as the absolute anti-commutativity of the conserved BRST-anti-BRST charges [cf. Eqs. (47), (48)] within the framework of the ACSA to BRST formalism. To achieve this goal, we note that, out of the equivalent expressions for the conserved BRST-anti-BRST charges presented in (45) and (46), one set of the conserved charges, ${\cal Q}_{(a)b}^{(1)}$, has been expressed in the BRST-anti-BRST exact forms.

With the identifications $s_b \leftrightarrow \partial_{\bar{\vartheta}}$ and $s_{ab} \leftrightarrow \partial_{\vartheta}$ in mind, we observe the following:
\begin{eqnarray}
{\cal Q}_b ^{(1)} & = & \frac{\partial}{\partial\bar{\vartheta}}\Big[\tilde N^{(b)}(t, \bar{\vartheta})\,\tilde N^{(b)}(t, \bar{\vartheta})\big\{{\dot{\bar{F}}}^{(b)}(t, \bar{\vartheta})\,F^{(b)}(t, \bar{\vartheta})-   {\bar{F}}^{(b)}(t, \bar{\vartheta})\,{\dot{F}}^{(b)}(t, \bar{\vartheta})\big\}\Big], \nonumber\\
&\equiv & \int d\bar\vartheta\,\Big[\tilde N^{(b)}(t, \bar{\vartheta})\,\tilde N^{(b)}(t, \bar{\vartheta})\big\{{\dot{\bar{F}}}^{(b)}(t, \bar{\vartheta})\,F^{(b)}(t, \bar{\vartheta})-   {\bar{F}}^{(b)}(t, \bar{\vartheta})\,{\dot{F}}^{(b)}(t, \bar{\vartheta})\big\}\Big], \nonumber\\
{\cal Q}_{ab} ^{(1)} & = & \frac{\partial}{\partial\vartheta}\Big[\tilde N^{(ab)}(t, \vartheta)\,\tilde N^{(ab)}(t, \vartheta)\big\{{\bar{F}}^{(ab)}(t, \vartheta)\,{\dot{F}}^{(ab)}(t, \vartheta) -   {\dot{\bar{F}}}^{(ab)}(t, \vartheta)\,F^{(ab)}(t, \vartheta)\big\}\Big], \nonumber\\
&\equiv & \int d\vartheta  \Big[\tilde N^{(ab)}(t, \vartheta) \tilde N^{(ab)}(t, \vartheta)\big\{{\bar{F}}^{(ab)}(t, \vartheta) {\dot{F}}^{(ab)}(t, \vartheta) - {\dot{\bar{F}}}^{(ab)}(t, \vartheta) F^{(ab)}(t, \vartheta)\big\}\Big].
\end{eqnarray}
As a consequence, it is straightforward to point out the  fact that we have the validity of the following relationships:
\begin{eqnarray}
\partial_\vartheta \; {\cal Q}^{(1)}_{ab} & = & 0 \quad \Longleftrightarrow \quad \partial_\vartheta^2 = 0 \quad \Longleftrightarrow \quad s_{ab}^2 = 0, \nonumber\\
\partial_{\bar\vartheta} \; {\cal Q}^{(1)}_{b} & = & 0 \quad \Longleftrightarrow \quad  \partial_{\bar\vartheta}^2 = 0 \quad \Longleftrightarrow \quad \, s_b^2 = 0.
\end{eqnarray}
In other words, we observe that the nilpotency, i.e., $\partial_\vartheta^2 = 0$ and $\partial_{\bar\vartheta}^2 = 0$, of the translational generators ($\partial_\vartheta, \partial_{\bar\vartheta}$) along the $(\vartheta)\bar{\vartheta}$-directions of the (1, 1)-dimensional chiral and anti-chiral super sub-manifolds (of the general (1, 2)-dimensional supermanifold) is responsible for capturing the off-shell nilpotency of the conserved BRST-anti-BRST charges ${\cal Q}_{(a)b}^{(1)}$.

\vskip 0.15 cm

To be more precise, we further note that the off-shell nilpotency of the conserved BRST charge ${\cal Q}_b^{(1)}$ is associated with the nilpotency, i.e., $\partial_{\bar{\vartheta}}^2 = 0$, of the translational generator $\partial_{\bar{\vartheta}}$ along the $\bar{\vartheta}$-direction of the (1, 1)-dimensional anti-chiral super sub-manifold. On the other hand, the off-shell nilpotency of the conserved anti-BRST charge ${\cal Q}_{ab}^{(1)}$ is intimately connected with the nilpotency, i.e., $\partial_{\vartheta}^2 = 0$, of the translational generator $\partial_{\vartheta}$ along the $\vartheta$-direction of the (1, 1)-dimensional chiral super sub-manifold of the general (1, 2)-dimensional supermanifold.

\vskip 0.15 cm

Finally, we focus on proving the absolute anti-commutativity [cf. Eq. (48)] of the conserved and nilpotent BRST-anti-BRST charges within the framework of the ACSA to BRST formalism. In this context, we note that, from the list of equivalent forms of conserved BRST-anti-BRST charges, the BRST charge ${\cal Q}_b^{(2)}$ has been expressed as an exact anti-BRST quantity. On the other hand, the conserved anti-BRST charge ${\cal Q}_{ab}^{(2)}$ has been written in the BRST exact form. With the identifications $s_b \longleftrightarrow \partial_{\bar{\vartheta}}$ and $s_{ab} \longleftrightarrow\partial_{\vartheta}$, we observe the sanctity of the following:
\begin{eqnarray}
&&{\cal Q}_b ^{(2)} =  \frac{\partial}{\partial\vartheta}
\Big[N^{(ab)}(t,\vartheta)\,\tilde N^{(ab)}(t,\vartheta)\,F^{(ab)}(t,\vartheta)\,{\dot{F}}^{(ab)}(t,\vartheta)\Big], 
\nonumber \\
&& \equiv  \int d\vartheta\,\Big[\tilde N^{(ab)}(t,\vartheta)\,\tilde N^{(ab)}(t,\vartheta)\,F^{(ab)}(t,\vartheta)\,{\dot{F}}^{(ab)}(t,\vartheta)\Big], \nonumber \\
&& {\cal Q}_{ab} ^{(2)}  =  \frac{\partial}{\partial\bar{\vartheta}}\Big[\tilde N^{(b)}(t,\bar{\vartheta})\,\tilde N^{(b)}(t, \bar{\vartheta})\,{\dot{\bar{F}}}^{(b)}(t, \bar{\vartheta})\,{\bar{F}}^{(b)}(t,\bar{\vartheta})\Big],\nonumber\\
&& \equiv  \int d\bar\vartheta\,\Big[\tilde N^{(b)}(t,\bar{\vartheta})\, \tilde N^{(b)}(t, \bar{\vartheta})\,{\dot{\bar{F}}}^{(b)}(t, \bar{\vartheta})\,{\bar{F}}^{(b)}(t,\bar{\vartheta})\Big].
\end{eqnarray}
As a consequence, it is straightforward that the following results are true, namely; 
\begin{eqnarray}
\partial_\vartheta \; {\cal Q}^{(2)}_{\cal B} & = & 0 \quad \Longleftrightarrow \quad \partial_\vartheta^2 = 0 \quad \Longleftrightarrow \quad s_{ab}^2 = 0, \nonumber\\
\partial_{\bar\vartheta} \; {\cal Q}^{(2)}_{ab} & = & 0 \quad \Longleftrightarrow \quad  \partial_{\bar\vartheta}^2 = 0 \quad \Longleftrightarrow \quad s_b^2 = 0.
\end{eqnarray}
Thus, it is clear that, in the ordinary space, the above equation is equivalent to equation (48), where we have proven the absolute anti-commutativity of the conserved and off-shell nilpotent BRST-anti-BRST charges. In the terminology of the ACSA to BRST formalism, we note that the absolute anti-commutativity between the BRST charge and the anti-BRST charge is deeply connected with the nilpotency, i.e., $\partial_{\vartheta}^2 = 0$, of the translational generator $\partial_\vartheta$ along the Grassmannian direction $\vartheta$ of the (1, 1)-dimensional chiral super sub-manifold of the general (1, 2)-dimensional supermanifold on which our 1D theory is generalized.

\vskip 0.15 cm

This should be contrasted with our earlier observation regarding the off-shell nilpotency of the BRST charge within the framework of ACSA to BRST formalism, where it is the nilpotency, i.e., $\partial_{\bar{\vartheta}}^2 = 0$, of the translational generator $\partial_{\bar{\vartheta}}$ along the Grassmannian direction $\bar{\vartheta}$ of the (1, 1)-dimensional anti-chiral super sub-manifold that plays a decisive role. 

\vskip 0.15 cm

Similar observations could be made for the proof of the absolute anti-commutativity between the anti-BRST charge and the BRST charge. However, for the sake of brevity, we refrain from making any such statements at this point in our discussion.\\


\section{Invariance of the Lagrangians: ACSA} 

\vskip 0.1 cm

We now capture the BRST-anti-BRST invariance of the coupled Lagrangians within the framework of the ACSA to BRST formalism and, thereby, prove the existence of the CF-type restriction (23) on our theory from the perspective of symmetry considerations. To be precise, we capture the BRST-anti-BRST invariance of the coupled  Lagrangians $L_{\cal B}$ and $L_{\bar{\cal B}}$ [cf. Eq. (40)]. Furthermore, we also express our observations from equation (43) in the language of the ACSA, which establishes the existence of the CF-type restriction: ${\cal B} + \bar{\cal B} + \dot{\bar{C}}\,C - \bar{C}\,\dot{C} = 0$ on our theory in terms of the invariance of the action integrals.

\vskip 0.15 cm

In this context, we first generalize the BRST-invariant Lagrangian $L_{\cal B}$ to its counterpart super Lagrangian $\tilde{L}_B^{(ac)}$ on the (1, 1)-dimensional anti-chiral super sub-manifold of the general (1, 2)-dimensional supermanifold (on which our theory is generalized) as follows:
\begin{eqnarray}
L_{\cal B}  &\longrightarrow &   \tilde L_{\cal B}^{(ac)}  =   P^{(ha)}_a (t, \bar\vartheta)\,{\dot A}^{(ha)} (t, \bar\vartheta) \nonumber\\  
& + & \frac{1}{2\,{A}^{(ha)} (t, \bar\vartheta)} \,  P^{(ha)}_a (t, \bar\vartheta)\,  P^{(ha)}_a (t, \bar\vartheta)\,\tilde N^{(b)} (t, \bar\vartheta)
\nonumber\\ 
& + &\frac {1}{2}\,\kappa\, \,\tilde N^{(b)}(t, \bar\vartheta)\, {\dot A}^{(ha)} (t, \bar\vartheta)  \nonumber\\
& + &2\,{\dot{\bar F}}^{(b)} (t, \bar\vartheta) \,F^{(b)}(t, \bar\vartheta) + {\bar F}^{(b)}(t, \bar\vartheta)\,{\dot F}^{(b)} (t, \bar\vartheta)\Big]
 \nonumber \\  
 & - &\tilde N^{(b)}(t, \bar\vartheta)\,{\dot{ \tilde N}}^{(b)} (t, \bar\vartheta)\,{\dot{\bar F}}^{(b)} (t, \bar\vartheta)\,F^{(b)} (t, \bar\vartheta) \nonumber \\ 
& - &\tilde N^{(b)}(t, \bar\vartheta)\,\tilde N^{(b)}(t, \bar\vartheta)\,{\dot{\bar F}}^{(b)} (t, \bar\vartheta)\,{\dot F}^{(b)} (t, \bar\vartheta) \nonumber \\ 
& - &{\bar F}^{(b)} (t, \bar\vartheta)\,{\dot{\bar F}}^{(b)} (t, \bar\vartheta)\,F^{(b)}(t, \bar\vartheta)\,{\dot F}^{(b)} (t, \bar\vartheta),
\end{eqnarray}
It can be noted that we have $\tilde{\cal B}^{(b)}(t, \bar{\vartheta}) = {\cal B} (t)$ due to the fact that $s_b\, {\cal B} = 0$. Therefore, although we have written $\tilde{\cal B}^{(b)}(t, \bar{\vartheta})$ in equation (55), it is an ordinary Nakanishi-Lautrup type auxiliary variable ${\cal B} (t)$ in our theory [cf. Eq. (51)].

\vskip 0.15 cm

Now, we are in a position to capture the BRST invariance of the Lagrangian $L_{\cal B}$ [cf. Eq. (40)] in the language of the ACSA as:
\begin{eqnarray}
\frac{\partial} {\partial \bar\vartheta }\, \tilde L_{\cal B}^{(ac)} & = & \frac {d}{dt}\Big [C\,L_f -  {\cal B}^2\,C - N^2\,{\cal B}\,\dot C 
- N\,\dot N\,{\cal B}\, C -  {\cal B}\,\bar C\,\dot C\, C \Big] \equiv   s_b\, L_{\cal B}.
\end{eqnarray}
Geometrically, this implies that the anti-chiral super Lagrangian $\tilde{L}_{\cal B}^{(ac)}$ is a combination of the appropriate (super)variables, such that its translation along the $\bar{\vartheta}$-direction of the (1, 1)-dimensional anti-chiral super sub-manifold results in a total ``time" derivative in the ordinary space. This, in turn, renders the action integral in the ordinary space invariant under the BRST symmetry transformations $s_b$ due to the Gauss divergence theorem.

\vskip 0.15 cm

It should be noted that the BRST transformations ($s_b$) are identified with the translational generator $\partial_{\bar{\vartheta}}$ [27-33] on the anti-chiral super sub-manifold of the general (1, 2)-dimensional supermanifold, on which our 1D system of FRW model is generalized.

\vskip 0.15 cm

We now discuss the anti-BRST invariance of our theory. To capture the anti-BRST symmetry invariance [cf. Eq. (40)] of the Lagrangian $L_{\bar{\cal B}}$, we first generalize the latter to the (1, 1)-dimensional chiral super sub-manifold of the general (1, 2)-dimensional supermanifold, on which our system of a 1D ordinary cosmological FRW model 
\begin{eqnarray}
L_{{\bar {\cal B}}} & \longrightarrow &   \tilde L_{{\bar {\cal B}}} ^{(c)}  =   P^{(hc)}_a (t, \vartheta)\,{\dot A}^{(hc)} (t, \vartheta) \nonumber\\  
& + & \frac{1}{2\,{A}^{(ha)} (t, \vartheta)} \,  P^{(hc)}_a (t, \vartheta)\,  P^{(ha)}_a (t, \vartheta)\,\tilde N^{(ab)} (t, \vartheta)
\nonumber\\ 
& + &\frac {1}{2}\,\kappa\, \,\tilde N^{(ab)}(t, \vartheta)\, {\dot A}^{(hc)} (t, \vartheta)  \nonumber\\
& + & 2\,{\bar F}^{(ab)}(t, \vartheta) \,{\dot F}^{(ab)}(t, \vartheta) + {\dot{\bar F}}^{(ab)}(t, \vartheta)\, F^{(ab)}(t, \vartheta)\big] \nonumber\\
& - & \tilde N^{(ab)}(t, \vartheta)\,{\dot {\tilde N}^{(ab)}}(t, \vartheta)\,{\bar F}^{(ab)}(t, \vartheta)\,{\dot F^{(ab)}}(t, \vartheta) \nonumber\\
& - & \tilde N^{(ab)}(t, \vartheta)\,\tilde N^{(ab)}(t, \vartheta)\,{\dot{\bar F}}^{(ab)}(t, \vartheta)\,{\dot F^{(ab)}}(t, \vartheta) \nonumber \\
 & - & {\bar F}^{(ab)}(t, \vartheta)\,{\dot{\bar F}}^{(ab)}(t, \vartheta)\,F^{(ab)}(t, \vartheta)\,{\dot F^{(ab)}}(t, \vartheta),
\end{eqnarray}
It can be noted that ${\tilde{\bar{\cal B}}}^{(ab)}(t, \vartheta) = \bar{\cal B}(t)$ due to the fact that $s_{ab} \, \bar{\cal B} = 0$. Thus, although we write ${\tilde{\bar{\cal B}}}^{(ab)}(t, \vartheta)$ in our chiral super Lagrangian $\tilde{L}_{\bar{\cal B}}^{(c)}$, it is actually an ordinary variable $\bar{\cal B}(t)$.

\vskip 0.15 cm

The anti-BRST invariance of the above chiral super Lagrangian can be expressed, in terms of the translational generator along the Grassmannian $\vartheta$-direction  as:
\begin{eqnarray}
\frac{\partial} {\partial \vartheta }\, \tilde L_{{\bar {\cal B}}}^{(c)} & = & \frac {d}{dt}\Big [ \bar C\,L_f -{{\bar {\cal B}}}^2\,\bar C + N^2\,{\bar {\cal B}}\,\dot {\bar C} 
+ N\,\dot N\,{\bar {\cal B}}\, \bar C  -  {\bar {\cal B}}\, \dot {\bar C} \,\bar C\,C \Big] \equiv  \, s_{ab}\, L_{{\bar {\cal B}}},
\end{eqnarray}
where $\partial_\vartheta$ is the translational generator [27-33] along the Grassmannian ($\vartheta$) direction of the (1, 1)-dimensional chiral super sub-manifold of the general (1, 2)-dimensional supermanifold.

\vskip 0.15 cm

Once again, geometrically, the chiral super Lagrangian $\tilde{L}_{\bar{\cal B}}^{(c)}$ is a specific combination of the appropriate chiral (super)variables such that its translation along the $\vartheta$-direction of the chiral super sub-manifold generates a total ``time" derivative (in the ordinary space), thereby rendering the action integral in the ordinary space invariant under the anti-BRST symmetry transformations ($s_{ab}$) due to the Gauss divergence theorem.
In the language of the ACSA to BRST formalism, we note that the super action integral $
\tilde{S}_1 = \int d\vartheta \int_{-\infty}^{+\infty} dt \, \tilde{L}_{\bar{\cal B}}^{(c)}$
remains invariant under the anti-BRST transformations.
\begin{eqnarray}
L_{{\bar {\cal B}}} & \longrightarrow & \tilde L_{{\bar {\cal B}}} ^{(ac)}  =    P^{(ha)}_a (t, \bar\vartheta)\,{\dot A}^{(ha)} (t, \bar\vartheta) \nonumber\\  
& + & \frac{1}{2\,{A}^{(ha)} (t, \bar\vartheta)} \,  P^{(ha)}_a (t, \bar\vartheta)\,  P^{(ha)}_a (t, \bar\vartheta)\,\tilde N^{(b)} (t, \bar\vartheta)
\nonumber\\ 
& + &\frac {1}{2}\,\kappa\, \,\tilde N^{(b)}(t, \bar\vartheta)\, {\dot A}^{(ha)} (t, \bar\vartheta)  \nonumber\\
& + & 2\,{\bar F}^{(b)}(t, \bar\vartheta) \,{\dot F}^{(b)}(t, \bar\vartheta) + {\dot{\bar F}}^{(b)}(t, \bar\vartheta)\,F^{(b)}(t, \bar\vartheta)\Big] \nonumber \\
 & - & \tilde N^{(b)}(t, \bar\vartheta)\,{\dot {\tilde N}}^{(b)}(t, \bar\vartheta)\,{\bar F}^{(b)}(t, \bar\vartheta)\,{\dot F}^{(b)}(t, \bar\vartheta)\nonumber\\
 & - & \tilde N^{(b)}(t, \bar\vartheta)\,\tilde N^{(b)}(t, \bar\vartheta)\,{\dot{\bar F}}^{(b)}(t, \bar\vartheta)\,{\dot F}^{(b)}(t, \bar\vartheta) \nonumber \\
& - & {\bar F}^{(b)}(t, \bar\vartheta)\,{\dot{\bar F}}^{(b)}(t, \bar\vartheta)\,F^{(b)}(t, \bar\vartheta)\,{\dot F}^{(b)}(t, \bar\vartheta).
\end{eqnarray}
In the above, it should be noted that we have generalized the perfectly anti-BRST invariant Lagrangian $L_{\bar{\cal B}}$ to its counterpart, the anti-chiral super Lagrangian $\tilde{L}_{\bar{\cal B}}^{(ac)}$, on the (1, 1)-dimensional anti-chiral super sub-manifold of the general (1, 2)-dimensional supermanifold. We are now in a position to apply a derivative $\partial_{\bar\vartheta}$ with respect to $\bar{\vartheta}$ on the above super Lagrangian, which yields the following expression (with the identification: $s_b \leftrightarrow \partial_{\bar\vartheta}$):
\begin{eqnarray}
\frac{\partial} {\partial \bar \vartheta}\; \tilde L_{{\bar {\cal B}}} ^{(ac)}  & = & 
\frac {d}{dt}\Big [ C\,L_f + N^2(\dot {\bar C}\, C\, \dot C + {\bar {\cal B}} \,  \dot C)  +    N\,\dot N\,(\bar C\, C\, \dot C + {\bar {\cal B}} \, C)\nonumber\\ 
& + &  \frac{d}{dt}\Big [{\cal B} +  {{\bar {\cal B}}} + \dot {\bar C}\, C - \bar C\, \dot C\Big]({\bar {\cal B}}\, C - N^2\, \dot C)\nonumber\\
&  - &  (2\,{\bar {\cal B}} - {\cal B})\, \bar C\, C\, \dot C -  {\bar {\cal B}} ^2\, C\Big] +   ({\cal B}+ {\bar {\cal B}} + \dot {\bar C}\, C - \bar C\, \dot C)\nonumber\\ 
&\times &(2\,{\bar {\cal B}}\,\dot C  -  2\,\dot {\bar C}\,C\,\dot C 
+  \bar C\, \ddot C \, C - N\,\dot N \, \dot C)
 \;\equiv \; s_b\, L_{{\bar {\cal B}}}.
\end{eqnarray}
The above equation leads to the derivation of the CF-type restriction in the sense that the anti-chiral super Lagrangian $\tilde{L}_{\bar{\cal B}}^{(ac)}$, when operated upon by $\partial_{\bar{\vartheta}}$, produces a total ``time'' derivative along with terms that vanish on the sub-manifolds of the space of variables defined by the CF-type restriction: ${\cal B} + \bar{\cal B} + \dot{\bar{C}}\,C - \bar{C}\, \dot{C}$. With the identification $s_b \leftrightarrow \partial_{\bar{\vartheta}}$, it is clear that we have obtained the same relationship as given in equation (43) in the ordinary space for the BRST symmetry transformation of $L_{\bar{\cal B}}$ (i.e., $s_b \,L_{\bar{\cal B}}$).

\vskip 0.15 cm

In exactly the same fashion, we can generalize the perfectly BRST invariant Lagrangian $L_{\cal B}$ to its counterpart chiral super Lagrangian $\tilde{L}_{\cal B}^{(c)}$ as follows:
\begin{eqnarray*}
L_{\cal B} & \longrightarrow &   \tilde L_{\cal B} ^{(c)}   =   P^{(hc)}_a (t, \vartheta)\,{\dot A}^{(hc)} (t, \vartheta) \nonumber\\  
& + & \frac{1}{2\,{A}^{(ha)} (t, \vartheta)} \,  P^{(hc)}_a (t, \vartheta)\,  P^{(ha)}_a (t, \vartheta)\,\tilde N^{(ab)} (t, \vartheta)
\nonumber\\ 
\end{eqnarray*}
\begin{eqnarray}
& + &\frac {1}{2}\,\kappa\, \,\tilde N^{(ab)}(t, \vartheta)\, {\dot A}^{(hc)} (t, \vartheta)  \nonumber\\
& + &2\, {\dot {\bar F}}^{(ab)}(t, \vartheta) \,{ F}^{(ab)}(t, \vartheta) 
+ {\bar F}^{(ab)}(t, \vartheta)\,{\dot F}^{(ab)}(t, \vartheta)\big] \nonumber\\
 & - &\tilde N^{(ab)}(t, \vartheta)\,{\dot {\tilde N}}^{(ab)}(t, \vartheta)\, {\dot {\bar F}}^{(ab)}(t, \vartheta)\,{F}^{(ab)}(t, \vartheta) \nonumber\\
 & - &\tilde N^{(ab)}(t, \vartheta)\,\tilde N^{(ab)}(t, \vartheta)\,{\dot{\bar F}}^{(ab)}(t, \vartheta)\,{\dot F}^{(ab)}(t, \vartheta) \nonumber \\
& - & {\bar F}^{(ab)}(t, \vartheta)\,{\dot{\bar F}}^{(ab)}(t, \vartheta)\,F^{(ab)}(t, \vartheta)\,{\dot F}^{(ab)}(t, \vartheta),
\end{eqnarray}
where all the symbols and notations have been clarified earlier. At this juncture, we apply a Grassmannian derivative $\partial_\vartheta$ on the above super Lagrangian, which yields the following:
\begin{eqnarray}
\frac{\partial} {\partial \vartheta}\; \tilde L_{\cal B} ^{(c)}  & = & 
 \frac {d}{dt}\Big [ \bar C\,L_f + N^2(\dot{\bar C}\,{\bar C}\, \dot C 
 - {\cal B} \,  \dot {\bar C}) +   N\,\dot N (\dot {\bar C} \,\bar C\, C - {\cal B} \, \bar C)\nonumber\\  
 & + & \frac{d}{dt}\Big [{\cal B} +  {{\bar {\cal B}}} + \dot {\bar C}\, C - \bar C\, \dot C\Big] ({\cal B}\, \bar C + N ^2\, \dot {\bar C})\nonumber\\
 & + &  (2\,{\cal B} - \bar{\cal B})\dot {\bar C}\, \bar C\, C - {\cal B} ^2\, \bar C\Big] +  ({\cal B}+ {\bar {\cal B}} + \dot {\bar C}\, C - \bar C\, \dot C)\nonumber\\ 
&\times & (2\,{\cal B}\,\dot {\bar C} + 2\,\dot {\bar C}\,\bar C\,\dot C +  \ddot{\bar C}\, \bar C \, C + N \,\dot N \, \dot {\bar C}) \;\equiv  \; s_{ab}\, L_{\cal B}.
\end{eqnarray}
Thus, we note that we have derived the observation made in equation (43). In other words, the ACSA to BRST formalism leads to the derivation of the CF-type restriction when we consider the anti-BRST invariance of the perfectly BRST invariant Lagrangian $L_{\cal B}$ as well as the BRST invariance of the perfectly anti-BRST invariant Lagrangian $L_{{\bar {\cal B}}}$ of our theory.\\

\section{Conclusions}

\vskip 0.5 cm

The cosmological Friedmann-Robertson-Walker (FRW) model provides a fundamental framework for understanding the large-scale structure and evolution of the universe under the assumptions of homogeneity and isotropy. Through the application of advanced symmetry principles, such as BRST-anti-BRST symmetry, the model offers deeper insights into re-parameterization (i.e. diffeomorphism) invariant dynamics, enriching our understanding of cosmological and quantum frameworks. This work demonstrates how quantum field theoretical approaches can be effectively integrated into classical cosmology, broadening the scope of the FRW model for quantum gravitational studies. These findings open avenues for further exploration in unifying general relativity (GR) with quantum mechanics, reaffirming the cosmological FRW model's centrality in both theoretical as well as observational cosmology.

\vskip 0.15 cm

In the current study, we have utilized the modified form of the supervariable approach, where the infinitesimal diffeomorphism transformation is incorporated methodically. Specifically, we first extended the transformation of the 1D infinitesimal diffeomorphism (re-parameterization), $t \rightarrow {t}' = t - \epsilon(t)$, to its superspace counterpart as defined in Eq. (11) on the (1, 2)-dimensional supermanifold, where the (anti-)ghost variables, $(\bar C)C$, appear as coefficients of the Grassmannian components.
This transformation, now viewed in superspace, was introduced into the supervariables defined on the (1, 2)-dimensional supermanifold. Afterwards, we proceeded to consider super expansions along all possible Grassmannian directions of the supermanifold. Only then did we apply the HC (Eq (16)) to derive the quantum BRST-anti-BRST transformations that correspond to the classical infinitesimal re-parameterization transformation: $t \rightarrow {t}' = t - \epsilon(t)$ in present cosmological FRW  model. We have labelled this method as the modified Bonora-Tonin supervariable approach (MBTSA).

\vskip 0.15 cm

One of the key outcomes of the present study is the derivation of the CF-type restriction: ${\cal B} + {\bar {\cal B}} + \dot{\bar C}C - \bar C \dot C = 0,$
which has been achieved by leveraging the power and potential of the modified BT-supervariable approach (MBTSA). This approach has not only facilitated the derivation of the BRST-anti-BRST symmetries for the target space variables, but it has also played a crucial role in deriving the BRST-anti-BRST symmetries for other variables within our theory. 
Furthermore, we have derived the BRST-anti-BRST symmetries for the remaining variables using the newly proposed ACSA to BRST formalism, where the BRST-anti-BRST invariant restrictions on the supervariables have been central to the analysis. The existence of the CF-type restrictions has been established through several key aspects of the theory: (i) the symmetry invariance of the coupled (but equivalent) Lagrangians in the ordinary space, (ii) the BRST-anti-BRST invariance of the super Lagrangians by exploiting the power of the ACSA to BRST formalism in superspace, and (iii) the proof of the absolute anti-commutativity of the Noether's conserved BRST-anti-BRST charges.

\vskip 0.15 cm

We have shown that the absolute anti-commutativity of the BRST-anti-BRST symmetries and their corresponding conserved charges, also the equivalence of the coupled  Lagrangians originate from the BRST-anti-BRST invariant CF-type restriction. 
In our present study, we have employed the modified BT-supervariable approach (MBTSA) to derive the nilpotent BRST-anti-BRST symmetry transformations for the variables \( a(t) \) and \( p_a (t) \) of the target space. The remaining BRST-anti-BRST symmetries for the other variables have been derived using the ACSA. One of the  {novel} insights of this work is the proof of the off-shell nilpotency and absolute anti-commutativity of the conserved BRST-anti-BRST charges within the framework of ACSA. A particularly interesting result is the observation that the absolute anti-commutativity of the BRST charge with the anti-BRST charge is deeply connected with the nilpotency (\( \partial_\vartheta^2 = 0 \)) of the translational generator \( \partial_\vartheta \) along the \( \vartheta \)-direction of the chiral super sub-manifold of the general (1, 2)-dimensional supermanifold. In contrast, the absolute anti-commutativity of the anti-BRST charge with the BRST charge is intimately related to the nilpotency (\( \partial_{\bar\vartheta}^2 = 0 \)) of the translational generator \( \partial_{\bar\vartheta} \) along the \( \bar\vartheta \)-direction of the  {anti-chiral} super sub-manifold of the general (1, 2)-dimensional supermanifold. 
Thus, in a certain sense, the ACSA formalism distinguishes between the  {chiral} and  {anti-chiral} super sub-manifolds in the context of proving the absolute anti-commutativity property.

\vskip 0.15 cm

The primary focus of our present work, however, is to perform the consistent BRST quantization of our 1D cosmological system by exploiting the full  {classical} infinitesimal re-parameterization symmetry transformations (9). This has been achieved with the aim that our findings will provide insights applicable to higher-dimensional diffeomorphism invariant theories (such as (super)strings and gravitational theories), where the methodology we present could prove useful. 
 It is gratifying to note that  the nature and form of the CF-type restriction we have derived is  {universal}, in the sense that we have obtained the same restriction in the BRST quantization of the 1D re-parameterization systems [36-40]

\vskip 0.15 cm 

Our future plan is to apply  standard techniques of the (anti-)chiral superfield approach (ACSA) to BRST formalism on the 
3D field-theoretic model [55] and various gauge invariant (i.e., BRST invariant) models for various theoretical and physical prospects.
 Our forthcoming objective is also to extend the BRST quantization to diffeomorphism invariant theories, such as (super)string theories, 
 gravitational theories, and cosmological theories,  which are at the forefront of research in theoretical high energy physics [56]. 
We are pleased to note that we have already made a modest contribution in this direction in our earlier and recent works [36-40], 
where we studied a two dimensional  diffeomorphism invariant model of the bosonic string [40] and derived the CF-type 
restrictions: ${\cal B}^a + {\bar {\cal B}}^a + i \left( C^b \,\partial_b \bar C^a + \bar C^b\, \partial_b C^a \right) = 0$
(where \( a, b = 0, 1 \)). These restrictions are the 2D version of the  {universal} CF-type restrictions for any D-dimensional diffeomorphism invariant theories [34, 35].\\

\section*{Acknowledgments}

We express our sincere gratitude to {\it Sunbeam Women's College, Varuna, Varanasi, India}, for providing a supportive research environment that contributed to the success of this study. We sincerely acknowledge the reviewer’s meticulous evaluation and insightful feedback, which have significantly enhanced the quality and depth of our manuscript.
\\

\vskip 0.7cm

\noindent
{\bf\large Data Availability}

\vskip 0.3cm

\noindent
No data were used to support this study.\\ 

\vskip 0.3cm 

\noindent
{\bf\large Conflicts of Interest}

\vskip 0.3cm

\noindent
The author declares that there are no conflicts of interest.\\ 

\vskip 0.5cm

\begin{center}
{\bf Appendix A: Lagrangian Formulation: Differential Gauge-Fixing Condition}\\
\end{center}

\vskip 0.2 cm
\noindent 
Gauge fixing is a crucial method in gauge theory, utilized to manage redundant degrees of freedom in field variables. It is essential for canonical quantization and removing extraneous gauge freedoms. The process involves imposing specific conditions to eliminate redundancy. These conditions must meet two fundamental criteria: (i) they should fully fix the gauge, leaving no residual freedom, and (ii) they should allow any configuration characterized by $N (t)$ and $a (t)$ to be transformed into one that satisfies the gauge condition.

In this context, we adopt the following gauge condition for the FRW model within an extended phase space framework [50, 51]:
\begin{equation}
\dot{N} = \frac{d}{dt} {\cal F} (a),
\end{equation}
where \( {\cal F}(a) \) is an arbitrary function dependent on \( a \).
This gauge condition has been scrutinized to ensure it provides a robust extended phase space formulation. Simply extending the phase space by adding gauge degrees of freedom is insufficient; the Lagrangian must also incorporate missing velocity terms. This is achieved through a differential gauge condition:
\begin{equation}
\dot{N} = \frac{d{\cal F}}{da} \dot{a}.
\end{equation}
To integrate this gauge condition into quantum theory, a corresponding gauge-fixing term is introduced in the invariant Lagrangian:
\begin{equation}
L_{gf} = {\cal B} \left(\dot{N} - \frac{d{\cal F}}{da} \dot{a}\right),
\end{equation}
where \( {\cal B} \) is a Lagrange multiplier that linearizes the gauge-fixing term.
In addition, the Faddeev-Popov ghost terms associated with this gauge-fixing condition for the FRW model are expressed as:
\begin{equation}
L_{gh} =  \bar C\, s_b (N - {\cal F}) = {\bar{C}} \left(\dot{N} - \frac{d{\cal F}}{da} \dot{a}\right) C + {\bar{C}} N \dot{C},
\end{equation}
where \( \bar{C} \) and \( C \) are ghost and anti-ghost variables with ghost numbers \( -1 \) and \( +1 \), respectively. These (anti-)ghost terms ensure the theory's consistency by removing non-physical degrees of freedom of the present cosmological FRW model of the universe.

The full extended Lagrangian, combining the original system, gauge-fixing, and ghost terms, is given by:
\begin{equation}
L_{ext} = p_a \dot{a} + \frac{1}{2a} p_a^2 N + \frac{1}{2} k N a + {\cal B} \left(\dot{N} - \frac{d{\cal F}}{da} \dot{a} \right) + {\bar{C}} \left(\dot{N} - \frac{d{\cal F}}{da} \dot{a} \right) C + {\bar{C}} N \dot{C}.
\end{equation}
 The off-shell nilpotent BRST-anti-BRST symmetry transformations for this extended Lagrangian $(L_{ext})$ are defined as:
\begin{eqnarray}
&& s_b N = (\dot{N} C + N \dot{C}), \;\; s_b a = \dot{a} C, \;\;  s_b p_a = \dot p_a\, C, \;\; s_b C = C \dot{C},\;\; s_b \bar{C} = {\cal B}, \;\;  s_b {\cal B} = 0, \nonumber\\
&& s_{ab} N = (\dot{N} \bar{C} + N \dot{\bar{C}}), \;s_{ab} a = \dot{a} \bar{C}, \; s_{ab} p_a = \dot p_a\, \bar C, \;s_{ab} \bar{C} = \bar{C} \dot{\bar{C}}, \;s_{ab} C = {\cal B}, \;  s_{ab} {\cal B} = 0.~~~
\end{eqnarray}
These transformations exhibit supersymmetry characteristics (bosonic variables transform into fermionic variables and vice versa), are nilpotent of order two (i.e., \( s_b^2 = 0 \) and \( s_{ab}^2 = 0 \)), and anti-commute with each other  (i.e., \( s_b s_{ab} + s_{ab} s_b = 0 \)).

The gauge-fixing and ghost terms are BRST-anti-BRST exact and expressed as:
\begin{equation}
L_{gf} + L_{gh} = s_b \, [\bar{C} \, (N - {\cal F})] \equiv s_{ab} \,[C \, (N - {\cal F})].
\end{equation}
Under these fermionic symmetry transformations, the extended Lagrangian transforms as a total time derivative:
\begin{align}
s_b L_{ext} &= \frac{d}{dt} \left[({\cal B}+ {\bar{C}} C) \left\{\left(\dot{N} - \frac{d{\cal F}}{da} \dot{a} \right) C + N \dot{C} \right\} \right], \\
s_{ab} L_{ext} & = -\,\frac{d}{dt} \left[({\cal B}+ \bar{C} \dot{C}) \left\{\left(\dot{N} - \frac{d{\cal F}}{da} \dot{a} \right) \bar{C} - N \,{\bar{C}} \right\} \right].
\end{align}
Consequently, the action integral (\( S = \int dt \, L_{ext} \)) for the cosmological FRW model remains invariant which confirms that the extended Lagrangian is BRST-anti-BRST invariant.\\


\begin{thebibliography}{99} 
\bibitem{BC1}    R. P. Feynman, {\it Acta Phys. Polon.} {\bf 24},  697 (1963)
\bibitem{BC2}    L. D. Faddeev,  V. N. Popov, {\it Phys. Lett.} B {\bf 25}, 29 (1967) 
\bibitem{BC3}    B. S. DeWitt, {\it Phys. Rev.} {\bf 160},  113 (1967); {\bf 162},  1195 (1967) 

\bibitem{BC4}     C. Becchi, A. Rouet, R. Stora, {\it Phys. Lett.} B {\bf 52},  344 (1974)   
\bibitem{BC5}     C. Becchi, A. Rouet, R. Stora, {\it Comm. Math. Phys.} {\bf 42},  127  (1975) 
\bibitem{BC6}     C. Becchi, A. Rouet, R. Stora, {\it Ann. Phys.}  (N. Y.) {\bf 98},  287 (1976)   
\bibitem{BC7}     I. V. Tyutin, {\it Lebedev Institute Preprint}, Report Number: {\bf FIAN-39} (1975)\\ (unpublished)
\bibitem{BC8}     L. Bonora, R. P. Malik, {\it Phys. Lett.} B {\bf 655}, 75 (2007)
\bibitem{BC9}     L. Bonora, R. P. Malik, {\it J. Phys.} A: {\it Math. Theor.} {\bf 43}, 375403 (2010)
\bibitem{BC10}     G. Curci, R. Ferrari, {\it Phys. Lett.} B {\bf 63}, 91 (1976)
\bibitem{BC11}    I. A. Batalin, G. A. Vilkovisky, {\it Phys. Lett.} B {\bf 102}, 27 (1981);\\                
                  {\it Phys. Rev.} D {\bf 28}, 2567 (1983);
                  {\it Nucl. Phys.} B {\bf 234}, 106 (1984) 
\bibitem{BC12}   I. A. Batalin, P. M. Lavrov,  I. V. Tyutin, {\it J. Math. Phys.} {\bf 31}, 1487 (1990)
\bibitem{BC13}   I. A. Batalin, P. M. Lavrov,  I. V. Tyutin, {\it J. Math. Phys.} {\bf 32}, 532 (1991)
\bibitem{BC14}   I. A. Batalin, P. M. Lavrov,  I. V. Tyutin, {\it J. Math. Phys.} {\bf 32}, 2513 (1991) 
\bibitem{BC15}     J. Thierry-Mieg, {\it J. Math. Phys.} {\bf 21}, 2834 (1980)
\bibitem{BC16}     M. Quiros, F. J. De Urries, J. Hoyos, M. L. Mazon, E. Rodrigues,\\
                   {\it J. Math. Phys.} {\bf 22}, 1767 (1981)              
\bibitem{BC17}    L. Bonora, M. Tonin, {\it Phys. Lett.} B {\bf 98}, 48 (1981) 
\bibitem{BC18}    L. Bonora,  P. Pasti,  M. Tonin, {\it Nuovo Cimento} A {\bf 64}, 307 (1981)
\bibitem{BC19}    L. Bonora,  P. Pasti,  M. Tonin,  {\it Annals of Physics} {\bf 144}, 15 (1982)
\bibitem{BC20}    R. Delbourgo, P. D. Jarvis, {\it J. Phys. A: Math. Gen.} {\bf 15}, 611 (1981)
\bibitem{BC21}    L. Baulieu, J. Thierry-Mieg, {\it Nucl. Phys.} B {\bf 197}, 477 (1982)
\bibitem{BC22}    L. Alvarez-Gaum e, L. Baulieu, {\it Nucl. Phys.} B {\bf 212}, 255 (1983) 
\bibitem{BC23}    R. P. Malik, {\it Eur. Phys. J.} C {\bf 60}, 457 (2009)
\bibitem{BC24}    R. P. Malik,  {\it J. Phys. A: Math. Theor.} {\bf 39}, 10575 (2006)
\bibitem{BC25}    R. P. Malik, {\it Eur. Phys. J.} C {\bf 51}, 169 (2007)
\bibitem{BC26}    R. P. Malik, {\it J. Phys.  A: Math. Theor.} {\bf 37}, 5261 (2004) 
\bibitem {BC27}   B. Chauhan, S. Kumar, {\it Adv. High Energy Phys.}  {\bf 2021}, 5518304  (2021)
\bibitem {BC28}  S. Kumar, B. Chauhan, A. Tripathi, R. P. Malik, \\ {\it Int. J. Mod. Phys.} A {\bf 37}, 2250003  (2022) 
\bibitem {BC29}  B. Chauhan, {\it Eur. Phys. J. Plus} {\bf 137}, 976 (2022) 
\bibitem {BC30}  B. Chauhan,  {\it Eur. Phys. Lett.} {\bf 140}, 40001 (2022) 
\bibitem {BC31}  A. Shukla, N. Srinivas, R. P. Malik, {\it Ann. Phys.} {\bf 394}, 98 (2018)
\bibitem {BC32}  B. Chauhan, S. Kumar, R. P. Malik, {\it Int. J. Mod. Phys.} A {\bf 37}, 2250003 (2022)
\bibitem {BC33}  B. Chauhan, T. Bhanja, {\bf arXiv: 2411.09948} [hep-th]
\bibitem {BC34}   L. Bonora, {\it Nucl. Phys.} B {\bf 912}, 103 (2016)
\bibitem {BC35}   L. Bonora, R. P. Malik, {\it Universe} {\bf 7}, 280 (2021)         
\bibitem {BC36}   B. Chauhan, A. Tripathi, A. K. Rao, R. P. Malik, 
                    \\ {\it Int. J. Mod. Phys. } A {\bf 37} 2250164 (2022) 
\bibitem {BC37}  A. Tripathi, B. Chauhan, A. K. Rao, R. P. Malik,\\
                    {\it Adv. High Energy Phys.}  {\bf 2020}, 2056629 (2020)
\bibitem {BC38}  A. Tripathi, B. Chauhan, A. K. Rao, R. P. Malik\\
                 {\it Adv. High Energy Phys.}  {\bf 2021}, 1236518 (2021)
\bibitem {BC39}  A. K. Rao, A. Tripathi, R. P. Malik, 
                 {\it Adv. High Energy Phys.}  {\bf 2021},  5593434 (2021) 
\bibitem {BC40}  A. K. Rao, B. Chauhan, R. P. Malik,  {\it Euro. Phys. J.}  Plus  {\bf 139}, 972 (2024)  
\bibitem {BC41}  G. Oyvind, H. Sigbjorn, {\it Einstein's General Theory of Relativity}, \\Springer, New York,  {\bf 195} (2007)
\bibitem{BC42}    B. S. Dewitt,  {\it Phys. Rev.}, {\bf 160} (1967) 1113. 
\bibitem{BC43}    D. L. Wiltshire,  {\it Cosmology: The Physics of the Universe},\\
                  {World Scientific}, Singapore (1996)
\bibitem{BC44}    A. Friedmann, {\it Zeit. f. Phys.}, {\bf 10}, 377 (1922)  
\bibitem{BC45}    A. Friedmann, {\it Zeit. f. Phys.}, {\bf 21}, 326 (1924)    
\bibitem{BC46}    H. P. Robertson,  {\it Astrophys. J.}, {\bf 82}, 284 (1935) 
\bibitem{BC47}    H. P. Robertson, {\it Astrophys. J.}, {\bf 83}, 187 (1935) 
\bibitem{BC48}    H. P. Robertson, {\it Astrophys. J.},  {\bf 83}, 257  (1936) 
\bibitem{BC49}    A. G. Walker, {\it Proc. Lond. Math. Soc. (2)},  {\bf 42},  90 (1937)
\bibitem{mon50}     S. Upadhyay,	{\it Prog. Theor.  Exper. Phys.},   {\bf 2015}, 093B06 (2015)
\bibitem{mon51}     S. Upadhyay, 	{\it Ann. Phys.},  {\bf 356}, 299 (2015)  
\bibitem{mon52}     F. Cianfrani, G. Montani, {\it Phys. Rev.} D,  {\bf  87}, 084025 (2013)
\bibitem{tp53}      T. P. Shestakova, {\it Class. Quantum Grav.},  {\bf 28}, 055009 (2011) 
\bibitem{BC54}    J. J. Halliwell, {\it Phys. Rev.} D.,  {\bf 38}, 2468 (1988) 
\bibitem{BC55}    R. Kumar, R. P. Malik, {\bf arXiv: 2412.10852} [hep-th]
\bibitem{BC56}    B. Chauhan, {\it et al.}, in preparation



\end{thebibliography}
\end{document}